\documentclass{CVM}

\usepackage{caption}
\usepackage{amsmath,lipsum}
\usepackage{cuted}

\usepackage{multicol,caption}
\usepackage{multirow}
\usepackage{longtable}
\usepackage{float}

\usepackage{colortbl}
\usepackage{arydshln}
\usepackage{multirow}
\usepackage{multicol}

\CVMsetup{
type      = {Review Article},
doi       = {s41095-0xx-xxxx-x},
title     = {Foveated Rendering: a State-of-the-Art Survey},
author    = { Lili Wang$^{1,2,3}$, Xuehuai Shi$^{1}$\cor{}, Yi Liu$^1$},
runauthor = {F. A. Author, S. B. Author, T. C. Author},
abstract  = {
  Recently, virtual reality (VR) technology has been widely used in medical, military, manufacturing, entertainment, and other fields.
  These applications must simulate different complex material surfaces, various dynamic objects, and complex physical phenomena, increasing the complexity of VR scenes. Current computing devices cannot efficiently render these complex scenes in real time, and delayed rendering makes the content observed by the user inconsistent with the user's interaction, causing discomfort. 
  Foveated rendering is a promising technique that can accelerate rendering. It takes advantage of human eyes' inherent features and renders different regions with different qualities without sacrificing perceived visual quality.
  Foveated rendering research has a history of 31 years and is mainly focused on solving the following three problems. 
  The first is to apply perceptual models of the human visual system into foveated rendering. The second is to render the image with different qualities according to foveation principles. The third is to integrate foveated rendering into existing rendering paradigms to improve rendering performance.
  In this survey, we review foveated rendering research from 1990 to 2021. 
  We first revisit the visual perceptual models related to foveated rendering. 
  Subsequently, we propose a new foveated rendering taxonomy and then classify and review the research on this basis. Finally, we discuss potential opportunities and open questions in the foveated rendering field.
  We anticipate that this survey will provide new researchers with a high-level overview of the state of the art in this field, furnish experts with up-to-date information and offer ideas alongside a framework to VR display software and hardware designers and engineers.
},
keywords  = {foveated rendering; virtual reality; real-time rendering},
copyright = {The Author(s)},
}

\begin{document}

\maketitle

    \begin{figure}[b] 
	\small\renewcommand\arraystretch{1.3}
	\begin{tabular}{p{80.5mm}} \toprule\\ \end{tabular}
	\vskip -4.5mm \noindent \setlength{\tabcolsep}{1pt}
	\begin{tabular}{p{3.5mm}p{80mm}}
			$1\quad $ & State Key Laboratory of Virtual Reality Technology and Systems, Beihang University, Beijing, 100000, China. Xuehuai Shi is the corresponding author. E-mail: wanglily@buaa.edu.cn, shixuehuaireal@buaa.edu.cn, rebreath@outlook.com.\\
			$2\quad $ & Peng Cheng Laboratory, Shengzhen, 518000, China.\\
			$3\quad $ & Beijing Advanced Innovation Center for Biomedical Engineering, Beihang University, Beijing, 100000, China.\\
		&\hspace{-5mm} Manuscript received: 2021-12-21; accepted: 2022-08-02\vspace{-2mm}
	\end{tabular} \vspace {-3mm}
\end{figure}


\section{Introduction}
\label{sec:introduction}

In recent years, virtual reality (VR) technology has been widely used in medical \cite{correa2009evaluation, hsieh2017vr, hsieh2018preliminary}, military \cite{rizzo2005human, lele2013virtual, ahir2020application}, manufacturing \cite{ong2013virtual, choi2015virtual, doolani2020review}, entertainment \cite{avila2014virtual, bialkova2017sound, saint2021survey}, and other fields \cite{puggioni2020content, ferdani20203d, tanenbaum2020make}. 
Despite the increasing computational power of devices, rendering overhead continues to increase owing to the diversification of surface materials of virtual objects, the increasing number of dynamic objects, and the higher complexity of physical phenomena to be simulated in VR applications. 
Moreover, Potter et al. \cite{potter2014detecting} demonstrated that the visual latency tolerance threshold for the human visual system (HVS) is approximately 13$ms$ \cite{potter2014detecting}, making it more difficult for these applications to meet  HVS real-time requirements.
If rendering results are too delayed, users will observe that the content is inconsistent with the interaction, which creates discomfort. Therefore, improving rendering performance is a critical factor in promoting the practicality of VR technology.

Foveated rendering is an accelerated rendering technology that allocates computing resources based on HVS perceptual models. More computing resources are allocated to the fovea of human eyes, while fewer are allocated to the periphery. The fovea is responsible for clear central vision because approximately half of the optic nerve fibers are distributed in the fovea of the retina, and the remaining half is distributed to the rest of the periphery \cite{hendrickson1984morphological}. Foveated rendering takes advantage of this inherent feature of human eyes. It performs different rendering qualities in different regions of the image. High-quality rendering is performed in the foveal region (fovea), and low-quality rendering is performed in the peripheral region (periphery). Therefore, foveated rendering can speed up rendering without sacrificing perceived visual quality.

Three challenges must be addressed in foveated rendering: the first is to use the perceptual model of the human visual system to guide foveated rendering, the second is to render different regions with different qualities, and the third is to integrate foveated rendering into existing rendering paradigms to improve rendering performance. 

\begin{itemize}
	\item \textbf{Using the perceptual model of the human visual system to guide foveated rendering.} This reduces computational overhead and ensures the user does not experience quality loss from the images generated. The basic idea of foveated rendering is to render the results of different qualities to different regions to accelerate the rendering process, therefore, it is first necessary to evaluate the rendering result quality based on the HVS.
	 A well-designed questionnaire for user studies is a straightforward approach to evaluate the visual quality of rendering results. However, this requires many user experiments to obtain effective results, which is extremely time-consuming.
	Prior to conducting large-scale user studies, researchers frequently use perceptual models and related metrics to evaluate the visual quality of rendering results and then perform user evaluations based on the results with satisfactory quality, thereby improving evaluation efficiency. 
	Visual quality is related to perceptual sensitivity\cite{loschky2001perceptual}. 
	The two most representative perceptual models related to the foveated rendering technique are the visual acuity and contrast sensitivity models.
	The visual acuity models describe the relationship between different regions in the visual field and the spatial resolution of the HVS.
	When applied to foveated rendering, the visual acuity models can be divided into the fall-off, binocular horopter, and ocular dominance models.
	Based on these models, foveated rendering allows low-quality rendering in regions with low spatial resolutions of the HVS and high-quality rendering in regions with high spatial resolutions to improve rendering performance without perceptual loss.
	The contrast sensitivity models describe the relationship between different contrast levels and the sensitivity of the HVS.
	The application of foveated rendering mainly includes various contrast sensitivity functions (CSFs), such as the spatial CSF, spatio-temporal CSF, spatio-luminance CSF, spatio-chromatic CSF, and critical flicker fusion.
	According to the contrast sensitivity model, foveated rendering can allocate less computational resources to the regions with low contrast sensitivity to improve rendering performance without losing visual perception.


	\item \textbf{Rendering different regions with different qualities.} Foveation principles should be considered to address this challenge. Level of detail (LoD) techniques in computer graphics provide a solution to render 3D scenes composed of geometric meshes with different qualities. This increases rendering efficiency by decreasing geometric mesh complexity and maintaining unnoticed visual quality reduction.    
	LoD techniques select different levels of details according to the viewpoint position and orientation. 
	When using LoD technology to render geometric meshes by foveated rendering \cite{luebke2001perceptually,zheng2018perceptual}, the user’s fovea is detected first, then the meshes that must be tessellated according to the fovea are finely controlled, finally refined meshes are used to generate high-quality rendering results in the foveal region.
	LoD technology is not only suitable for geometric meshes but also for other data representations, such as point cloud data \cite{schutz2019real}. 
	
	In addition to the degree of mesh tessellation, the rendering sampling rate in rendering is also an essential factor that directly affects the quality of the resulting image. 
	User behavior and performance have been evaluated in user studies\cite{loschky2000user,loschky2001perceptual,parkhurst2002variable}. The results showed that users could not distinguish images with a reduced sampling rate below the perceptual thresholds in the peripheral regions from full resolution images. 
	Multi-spatial resolutions based foveated rendering methods perform high-resolution sampling for foveal regions and some important regions that users may notice, and low-resolution sampling for peripheral regions.
	Alongside the concept of multi-spatial resolution, multi-temporal, multi-luminance, and multi-color resolution can also be used to accelerate foveated rendering.
	In this survey, these foveation principles are essential factors in our taxonomy of foveated rendering technologies.


	%

	\item \textbf{Integrating foveated rendering into existing rendering paradigms to improve rendering performance.}
	Rasterization is the most widely studied rendering paradigm in foveated rendering \cite{duchowski2014reducing, turner2018phase}. 
	To rasterize the image with different resolutions in screen space, early research first rasterized the full resolution image and then reduced the image resolution in the desired region with time-consuming filters, which  opposed the goal of foveated rendering. 
	Since 2012, foveated rendering using rasterization has only performed high-resolution rendering in foveal regions and some important regions that users may notice, and low-resolution rendering in peripheral regions \cite{guenter2012foveated,bastani2020smoothly,stengel2016adaptive,tursun2019luminance,tavakoli2019scene}.
	Because implementing rasterization into foveated rendering may create multiple rendering passes, a general rendering pipeline to rasterize pixels with foveated rendering in a single render pass was introduced, thereby further improving rendering efficiency \cite{parkhurst2002variable}. 
	The  ray tracing rendering paradigm can control the number of rays emitted by each pixel. This directly supports the multi-spatial resolution. Therefore, many researchers implemented this approach with foveated rendering \cite{koskela2016foveated,molenaar2018towards,koskela2019foveated,koskela2020foveated}.
	Besides rasterization and ray tracing, some studies focus on implementing other rendering paradigms into foveated rendering, such as ray casting, instant radiosity, and  neural rendering  \cite{levoy1990gaze, wang2020foveated, bruder2019voronoi, kaplanyan2019deepfovea}.
	Hence, the rendering paradigm is also an essential factor in our taxonomy.
\end{itemize}

This survey aims to review the state-of-the-art in the field of foveated rendering, and to discuss foveated rendering methods with different input data types, foveation principles, and rendering paradigms in design and implementation, especially 3D foveated rendering methods that emerged in the past 10 years.

This section briefly introduced foveated rendering  concepts and challenges.
Section \ref{sec_VisualPerceptionTheory} discusses the application of HVS perceptual models to foveated rendering. Section \ref{sec_3DFRT} proposes foveated rendering method taxonomies and classifies previous methods. Section \ref{sec_earlyWork} revisits early related foveated rendering research from 1990 to 2011 based on the taxonomy. Section \ref{sec_recentWork} reviews methods that emerged over the past decade based on the taxonomy. Research conducted in the first 20 years and the last 10 years are separated because the focus of foveated rendering research has changed.
Finally, Section \ref{sec_OOQ} discusses foveated rendering open questions and opportunities.

\section{Applying Visual Perceptual Models in Foveated Rendering} \label{sec_VisualPerceptionTheory}

First, the HVS visual features involved in foveated rendering are briefly summarized. Then, perceptual models are introduced after which we discuss the application of these models in foveated rendering. We recommend Weier's survey \cite{weier2017perception} to those who wish to establish a more comprehensive understanding of perception-based rendering techniques.

\subsection{HVS Features involved in Foveated Rendering}
Currently, HVS primary visual features involved in foveated rendering include visual acuity and contrast sensitivity. Both are described as follows.

\subsubsection{Visual Acuity}
Visual acuity refers to the ability to discern shapes and details of objects \cite{cline1980dictionary}. As the main HVS feature widely used in foveated rendering, it has the following properties:

\begin{figure*}[htbp] 
	\setlength{\abovecaptionskip}{5pt}   
	\setlength{\belowcaptionskip}{-10pt}   
	\centering
	\includegraphics[width=\linewidth]{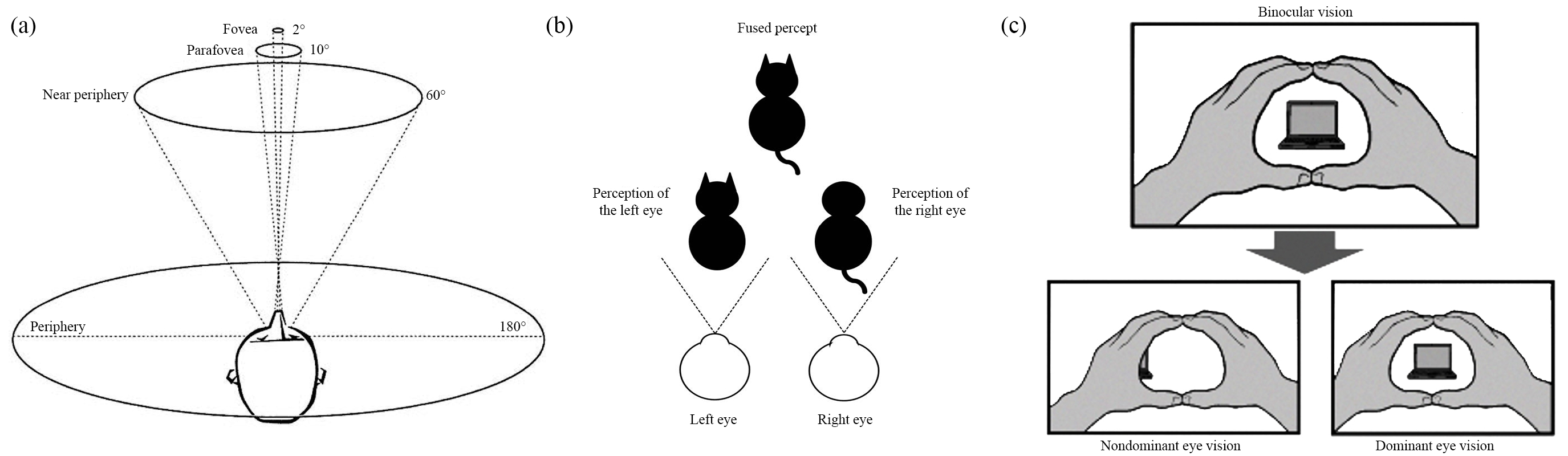}
	\caption{
		{\leftskip=0pt \rightskip=0pt plus 0cm
			(a) Schematic illustration of foveal$\backslash$peripheral vision from Ivanvcic et al. \cite{ivanvcic2019impact}. 
			The eccentric angle of the foveal region is very small, while the eccentric angle of the parafoveal region is up to 10$^\circ$. The eccentric angle of the near peripheral region is about 60$^\circ$ and that of the peripheral region area is 180$^\circ$. (b) Schematic illustration of fusional vision from Schaadt et al. \cite{schaadt2015disorders}. Two laterally placed eyes provide us two horizontally shifted and disparate images of the visual scene, which are continuously integrated into a single percept. (c)  Schematic illustration of 
			dominant eye. Compared with the nondominant eye, the dominant eye contributes more to the binocular vision.
		}
	}
	\label{fig:vap}
\end{figure*}

\begin{itemize}
	\item \textit{Foveal$\backslash$Peripheral Vision.} Human visual acuity is not uniform over the whole visual field. When a person looks at an object, the foveal vision scene details can be recognized, however,  the peripheral vision scene cannot be clearly recognized \cite{strasburger2011peripheral}.
	That is, the HVS has higher visual acuity in the fovea of the human visual region and is the basis of foveated rendering. Figure \ref{fig:vap} (a) shows a schematic illustration of the foveal$\backslash$peripheral vision.
	
	\item \textit{Fusional Vision.} The movement of both eyes enables the fusion of monocular images producing binocular vision. In fusional vision, the area where objects are perceived as single unified objects when viewed with both eyes is called Panum's fusional area \cite{fender1967extension}. The scene out of the Panum's fusional area is recognized as a ``double image" with lower image quality and less visual realism \cite{georgeson2014binocular}. It can be used to simplify the scene out of Panum's fusional area for efficient foveated rendering. Figure \ref{fig:vap} (b) shows a schematic illustration of fusional vision.  
	
	\item \textit{Dominant Eye.} Both eyes have different sensitivity to visual stimuli in the HVS, i.e., one is more sensitive than the other, and the eye with higher sensitivity is called the dominant eye \cite{porac1976dominant}. 
	Less computational resources can be allocated for the non-dominant eye to speed up rendering when performing foveated rendering for binoculars. Figure \ref{fig:vap} (c) shows a schematic illustration of the dominant eye.
	
\end{itemize}

\subsubsection{Contrast Sensitivity}
Contrast sensitivity refers to the ability to distinguish between foreground objects and background \cite{robson1966spatial}. This varies from individual to individual, reaching a maximum at approximately the age of 20, and subsequently decreases with age. Other factors (such as cataracts and diabetic retinopathy) can also cause a decrease in contrast sensitivity. Contrast sensitivity can be considered from the following distinct aspects:

\begin{itemize}
	\item \textit{Spatial Contrast Sensitivity.} This refers to the HVS sensitivity in recognizing patterns at different frequencies \cite{campbell1968application}.
	For certain scene regions to be rendered, where the HVS frequency is less sensitive, it is possible to perform lower quality rendering in these regions to improve efficiency.

	\item \textit{Spatio-temporal Contrast Sensitivity.} This refers to the HVS spatial contrast sensitivity at different retinal velocities \cite{kelly1979motion}. 
	Rendering quality can be dynamically adjusted to the current retinal velocity to improve 
foveated rendering quality and performance.
	
	\item \textit{Spatio-luminance Contrast Sensitivity.} This refers to the HVS spatial contrast sensitivity at different luminances.
	Adjusting rendering quality according this aspect in environments with different luminances can also improve foveated rendering quality and performance.
	
	\item \textit{Spatio-chromatic Contrast Sensitivity.} This refers to the HVS contrast sensitivity through grating stimulation with sinusoidally changing colors \cite{mullen1985contrast}.
	In particular environments, such as bars, fog, and other scenes with prominent theme colors, foveated rendering can also use this aspect to improve perceptual quality.
	
\end{itemize}

\begin{figure}[htbp] 
	\centering
	\includegraphics[width=\linewidth]{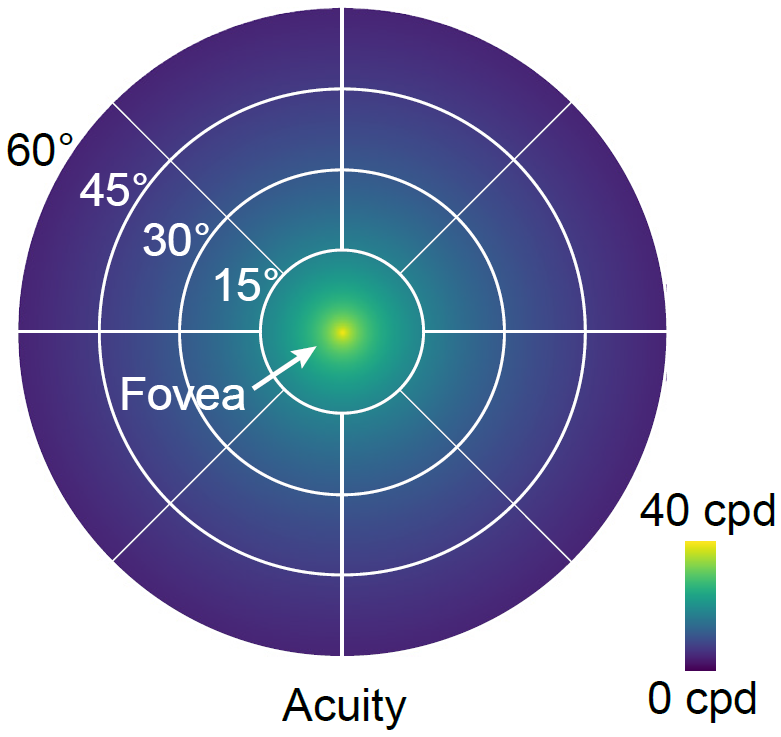}
	\caption{
		{\leftskip=0pt \rightskip=0pt plus 0cm
			The visual acuity fall-off model proposed by Geisler et al. \cite{geisler1998real}. 
			The HVS visual acuity is the largest (40$cpd$) when the eccentricity angle is 0$^{\circ}$. With the increase of eccentricity, visual acuity decreases linearly. When the eccentricity angle exceeds 45$^{\circ}$, visual acuity decreases to 0$cpd$. Image from Krajancich et al. \cite{krajancich2021perceptual}. 
		}
	}
	\label{fig:vafm}
\end{figure}

\subsection{Perceptual Models and Foveated Rendering Applications}
Based on the HVS features, perceptual models were proposed and used in foveated rendering to approximate the HVS functions and features through mathematical descriptions. These models could guide foveated rendering design and determine the perceptual quality of the rendering result. This section reviews the perceptual models, and their application in foveated rendering based on the HVS features discussed. 

\subsubsection{Visual Acuity Models}
Visual acuity models describe the function of visual acuity with neural and optical factors. Various visual acuity models have been developed based on \textit{foveal$\backslash$peripheral vision}, \textit{fusional vision} and the \textit{dominant eye}.

\textit{Visual Acuity Fall-off Model.} This is the psychophysical model that shows the degradation behavior of visual acuity with eccentricity \cite{weymouth1958visual}.
Weymouth et al. \cite{weymouth1963visual} demonstrated that acuity could be measured in terms of MAR (minimum angular resolution). A linear model matches both anatomical data and performance results on many vision tasks. 
Daniel et al. \cite{daniel1961representation} proposed the cortical magnification factor (CMF), which provides the mapping from the visual angle to a cortical diameter in millimeters. The magnification factor is the largest in 0-20$^{\circ}$ and decreases with eccentricity for the periphery.
Levi et al. \cite{levi1985vernier} stated that MAR increases linearly with eccentricity in the first 20-30$^{\circ}$. The higher the eccentricity, the faster the angular dimension rises.
From the center of the visual vision to the peripheral vision, the spatial sensitivity is reduced by 35$\times$ \cite{nakayama1990properties}. Figure \ref{fig:vafm} shows an example of the visual acuity fall-off model from Geisler et al. \cite{geisler1998real}.

In early foveated rendering research,
Levoy et al. \cite{levoy1990gaze} combined the ray casting method used for volume rendering with the visual acuity fall-off model. For each pixel on the image plane, they first calculated the eccentricity of the pixel, then obtained the visual acuity of this pixel based on the eccentricity and the visual acuity fall-off model, and finally modulated the number of rays casting on this pixel and the number of samples per unit length of each ray based on acuity to generate the rendering result.
Some studies combined vertex decimation algorithms with the visual acuity fall-off model to dynamically adjust the number of vertices extracted based on the visual sensitivity corresponding to each pixel to achieve LoD, i.e., higher accuracy for face slices in the gaze point region and lower accuracy for face slices in the surrounding region \cite{ohshima1996gaze, luebke2000perceptually, luebke2001perceptually, parkhurst2001evaluating}. The behavioral performance cost of a series of perceptual experimental surface gaze level-of-detail techniques can be offset by the behavioral performance gain from increased rendering speed.
In more recent research on the topic, 
Gunter et al. \cite{guenter2012foveated} simulated the acuity drop by rendering three nested layers of increasing angular diameters and decreasing resolution around the gaze direction. These layers were fused into the final result image. This work employed the CMF to decrease resolution, achieved significant shading reductions, and introduced overhead by repeating rasterization. 
Vaidyanathan et al. \cite{vaidyanathan2014coarse} proposed an architecture for the flexible control of shading rates in a GPU pipeline and tested their architecture for foveated rendering with a simplified visual acuity fall-off model. 
Weier et al. \cite{weier2016foveated} combined the visual acuity fall-off model with the re-projection technique and applied it in the ray tracing algorithm for head mounted displays (HMDs). For each frame, if the re-projection technique cannot reuse the rendering result of the previous frame, the number of sampling rays required for the current pixel is determined by the corresponding visual acuity, with higher visual acuity requiring a larger number of sampling rays and lower visual acuity requiring a smaller number.

\begin{figure}[htbp] 
	\centering
	\includegraphics[width=\linewidth]{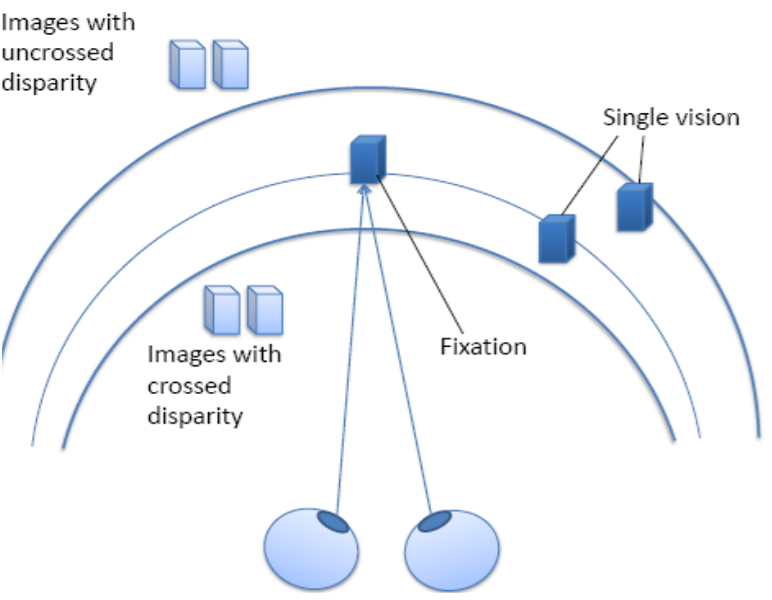}
	\caption{
		{\leftskip=0pt \rightskip=0pt plus 0cm
			Panum's Fusional Area: objects within the area are perceived as single images, objects further away are seen perceived with uncrossed disparity, and the objects closer to the viewer with crossed disparity. Image from Mikkola et al. \cite{mikkola2010relative}. 
		}
	}
	\label{fig:fusion}
\end{figure}

\textit{Binocular Horopter Model.} An empirical binocular horopter model was introduced by Panum et al. \cite{panum1858physiologische}, which reported that the sensory mechanism of the HVS fuses the images perceived by two eyes. This fusion leads to a single vision experience in the average visual direction and Panum's fusional area, as shown in Figure \ref{fig:fusion}. Mitchell et al. \cite{mitchell1966review} measured the upper limit of the parallax range, which was used to represent the upper disparity tolerance of the sensory mechanism for fusion.

In foveated rendering, Ohshima et al. \cite{ohshima1996gaze} used fusion vision theory to control the geometric meshes level of detail. They reduced the complexity of geometry out of the fusional area to accelerate rendering. 
Based on the theory of fusion vision, many other studies focus on improving the depth-of-field blur effects \cite{hillaire2008using, mantiuk2011gaze, duchowski2014reducing, 2014Gaze, mauderer2014depth, kusha2016Gaze, weier2018foveated, kang2020depth}.

\textit{Ocular Dominance Model.}
The ocular dominance model was proposed by Banister et al. \cite {porac1976dominant}, which showed that the HVS tends to use one eye instead of both to perceive the scene. Shneor et al. \cite {shneor2006eye} evaluated the effect of ocular dominance under non-rivalry conditions and concluded that the dominant eye has priority in visual processing and may inhibit the performance of the non-dominant eye. Ko{\c{c}}tekin et al. \cite{kocctekin2013relation} evaluated the performance of the dominant eye for color vision discrimination ability among medical students with normal color vision and concluded that the dominant eye takes priority in the r/g color spectral region, probably including inhibition of the non-dominant eye. 

Meng et al. \cite{meng2020eye} adopted the ocular dominance model into foveated rendering and rendered the non-dominant display with a more aggressive foveation to accelerate foveated rendering on HMDs.

\subsubsection{Contrast Sensitivity Models}
In foveated rendering, contrast sensitivity models mainly describe the HVS ability to distinguish objects from the background behind objects at different spatial frequencies \cite{owsley2003contrast}, such as contrast sensitivity functions (CSFs); and the threshold at which an identical flickering stimulus varies in percept from flickering to stable, such as critical flicker fusions (CFFs).
In CSF research, attention is paid not only to the influence of the most fundamental spatial frequencies, but also the influence of temporal frequencies, luminances, and colors \cite{krajancich2021perceptual, 2013Measurements, chwesiuk2019measurements}.
In CFF research, attention is focused on measuring the threshold at which the HVS can perceive the stable flickering stimulus in the temporal domain \cite{tyler1990analysis, krajancich2021perceptual}.

\begin{figure*}[htbp]  
	\centering
	\includegraphics[width=\linewidth]{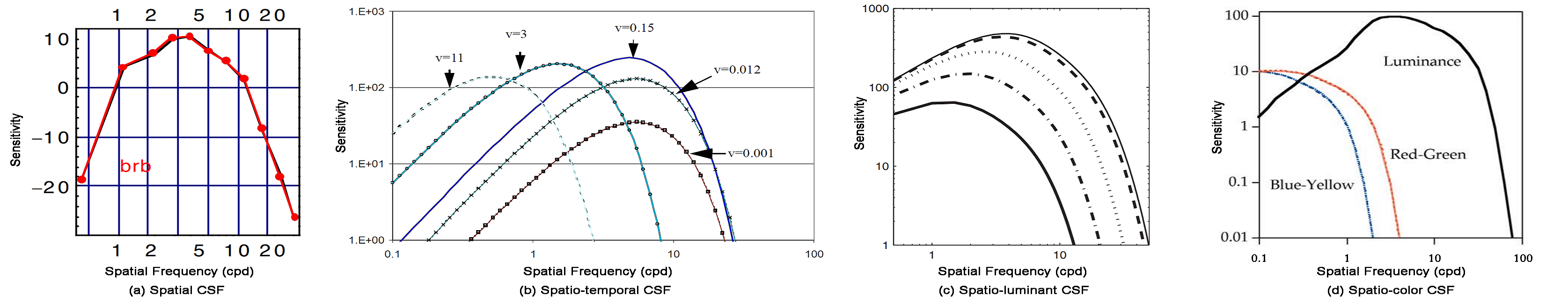}
	\caption{
		{\leftskip=0pt \rightskip=0pt plus 0cm
			CSFs over the range of spatial frequencies, temporal frequencies, luminances, and colors. (a) Spatial CSF with the measured data from Watson et al. \cite{watson2000vis}. (b) Spatio-temporal CSF derived from sensitivity measurements in Yee et al. \cite{yee2001spatiotemporal}, where $v$ is the velocities of the retinal images that measured in $d/s$. (c) Spatio-luminance CSF measured by Barten's model in  Westland et al.  \cite{westland2006model} for stimulus of size 10 $cpd$ and mean luminance 50 (thin solid line), 25 (dashed line), 2.5 (dotted line), 0.25 (dashdotted line), and 0.025 (thick solid line) $cd/m^2$. (d) Spatio-chromatic CSF for black-white, red–green and yellow–blue contrast from Fairchild et al. \cite{fairchild2013color}.
		}
	}
	\label{fig:CSFs}
\end{figure*}

\textit{Spatial CSF.} This was first proposed by Schade et al. \cite{schade1956optical} and measured the contrast detection threshold of the most sensitive part of the range in a logarithmic scale range, and distributed evenly on the most sensitive part of this range, typically 1-16$cpd$ (cycles per degree).
Nowadays, the most commonly used spatial CSF is the threshold set measured by Watson et al. \cite{watson2000vis} as a function of spatial frequency. 
Examples in Figure \ref{fig:CSFs} (a) show that spatial CSF peaks between 4-5$cpd$ and falls rapidly at higher frequencies. 

In early foveated rendering research, many researchers used spatial CSF to accelerate rendering by reducing the geometry complexity of the scene in the high static spatial frequency region\cite{xia1996dynamic, hoppe1997view, luebke1997view}. Because the HVS is less sensitive to high-frequency patterns in the peripheral regions, the HVS can tolerate greater errors in the high-frequency regions of the rendered scene. 
Recently, Patney et al. \cite{patney2016towards} introduced a novel anti-aliasing algorithm to help recover peripheral details that are resolvable by our eyes. This algorithm provides details that the periphery of the HVS can perceive. Koskela et al. \cite{koskela2016foveated} demonstrated that the smallest detail that humans can resolve is 60$cpd$ on average. If a rendering system could be built capable of showing 60$cpd$, 95\% of the rendered detail would be excessive. Then Koskela et al. \cite{koskela2019foveated} proposed a novel Visual-Polar coordinate space and distributed the samples according to the spatial CSF in the Visual-Polar coordinate space.

\textit{Spatio-temporal CSF.} 
The HVS contrast sensitivity not only changes with spatial frequency but also with retinal velocities. 
The spatio-temporal CSF measures the HVS contrast sensitivity with spatial frequency and retinal-image motion. Kelly et al. \cite{kelly1979motion} measured the CSF by allowing the user to observe sine waves with different retinal velocities. Contrast sensitivity varies significantly with retinal velocity. 
Liu et al. \cite{1984Contrast}, and Flipse et al. \cite{1988Contrast} reported that if the velocity of the retinal image is identical, the contrast sensitivity of the eye during fixation and pursuit will be equal, i.e., the motion of the retinal image, not the motion of the eye, determines contrast sensitivity. Figure \ref{fig:CSFs} (b) shows that the temporal CSF varies with different velocities of the retinal images.

In foveated rendering, Yee et al. \cite{yee2001spatiotemporal} constructed a spatio-temporal error tolerance map based on a spatio-temporal CSF to accelerate rendering and achieved a significant improvement in speed. Stengel et al. \cite{stengel2016adaptive} introduced a sampling scheme combined with a spatio-temporal CSF, which performs shading on regions of essential features in the image, and interpolates the remaining regions, to avoid affecting user perceived quality.

\textit{Spatio-luminance CSF.} The HVS contrast sensitivity changes with luminance under the same spatial frequency. 
Meeteren et al. \cite{1972Resolution} measured contrast sensitivity with a luminance ranging from 0.0001 to 10$cd/m^2$ in the case of a spatial frequency ranging from 0.5 to 30$cpd$. 
Kim et al. \cite{2013Measurements} extended the contrast sensitivity measure into higher luminance levels (150$cd/m^2$) with lower spatial frequencies, down to 0.125$cpd$. 
Higher luminance levels are more relevant to photopic vision, and low frequencies are required to observe and model the CSF band-pass characteristic, especially for low luminance levels. Figure \ref{fig:CSFs} (c) shows that the luminance CSF varies with different mean luminances.

In foveated rendering, Stengel et al. \cite{stengel2016adaptive} proposed a luminance map to adjust the sampling probability such that the number of colored samples is further distributed in the image with essential features.
Tursun et al. \cite{tursun2019luminance} proposed a new luminance-contrast-aware foveated rendering technique, which analyzed the local luminance contrast of the image to obtain a particular foveation to improve computational savings.

\textit{Spatio-chromatic CSF.} The HVS contrast sensitivity changes significantly with sinusoidally changing colors at the same spatial frequency. 
Mullen et al. \cite{1991Colour} performed experiments that compare the decline in contrast sensitivity between the color-only (red-green) gratings and the monochromatic luminance gratings in the entire field of view when the spatial frequency is 2$cpd$, at the center of the fovea and the eccentricity are 10$^\circ$ and 18$^\circ$. 
Anderson et al. \cite{anderson1991human} measured the CSF for eccentricities from 0$^\circ$ to 55$^\circ$ for chromatic red-green sinusoidal stimuli and reported that chromatic contrast declines more steeply than luminance contrast with eccentricity. Mullen et al. \cite{2002Differential} measured the cone contrast sensitivities for sine-wave grating stimuli (smoothly enveloped in space and time) for two colors (red-green and blue-yellow) and monochromatic luminance at a range of eccentricities in the nasal field (0-25$^\circ$).	They identified that red-green cone opponency has a steep decline away from the fovea, while the loss in blue-yellow cone opponency is more gradual, showing a similar loss to that found for achromatic vision. 
Mullen et al. \cite{2005Does} measured the cone contrast for red-green and blue-yellow colors. The results showed that red-green cone opponency declines steeply across the human periphery and becomes behaviorally absent by 25-30$^\circ$. 
Chwesiuk et al. \cite{chwesiuk2019measurements} reported that the color directions closer to the chromatic green-to-red axis show higher contrast sensitivity in comparison with achromatic stimuli, while for the yellow-to-blue axis, the sensitivity is lower. Figure \ref{fig:CSFs} (d) shows that the color CSF varies with black-white, red–green and yellow–blue.

Duchowski et al. \cite{duchowski2007foveated} introduced the possibility of developing a perceptually-based color degradation metric, which can be used to accelerate foveated rendering. They also investigated the peripheral color reduction with the color CSF, the results suggested that peripheral chromaticity cannot be reduced within the central 20$^\circ$ visual angle.

\textit{Critical Flicker Fusion.}
Besides CSFs that focus on distinguishing objects from the background, researchers measured the threshold at which an identical flickering stimulus varies in percept from flickering to stable, i.e., critical flicker fusion (CFF) \cite{tyler1990analysis, krajancich2021perceptual}.
Tyler et al. \cite{tyler1990analysis} introduced the Ferry-Porter law considering spatio-temporal frequency and luminance, which described that CFF increases linearly with log retinal luminance and log stimulus area, respectively. Tyler et al. \cite{tyler1993eccentricity} showed that the Ferry-Porter law also extends to higher eccentricities.
Krajancich et al. \cite{krajancich2021perceptual} introduced a model to measure the eccentricity-dependent critical flicker fusion thresholds for space, time, and luminance. This showed that the CFF varies with spatial frequency and luminance and exhibited an anti-foveated effect, with the highest thresholds observed in the near–mid periphery of the visual field.
Although no research directly applied the eccentricity-based CFF to foveated rendering algorithms, this provided a new model to improve foveated rendering efficiency.

\section{3D Foveated Rendering Taxonomies} \label{sec_3DFRT}

Recent surveys proposed several taxonomies to classify existing foveated rendering techniques. 
In Weier's survey \cite{weier2017perception}, the authors referred to foveated rendering methods as measurement-based perceptual approaches and classified them into two catalogs: one based on scene simplification, the other based on adaptive sampling. The methods in the first catalog are object-space methods. They use geometry techniques, such as LoD, to significantly reduce the scene's complexity using the visual acuity model or CSF according to the user's gaze position, thereby significantly improving time performance. The methods in the adaptive sampling class adaptively calculate the sampling rate in rendering paradigms, such as rasterization or ray tracing based on the visual acuity model or CSF.

\begin{table*}[bhpt]
	\footnotesize
	\centering
	\caption{The classification matrix is produced by combining RDF classification (letters) with motion classification (numbers) from Spjut et al. \cite{spjut2020toward}.}
	\label{tab:2020toward}
	
	\begin{tabular}{|m{1.8cm}<{\centering}|m{3.3cm}<{\centering}|m{3.3cm}<{\centering}|m{3.3cm}<{\centering}|m{3.3cm}<{\centering}|}
		\cline{2-5}
		\multicolumn{1}{c|}{} & \textbf{Class A} & \textbf{Class B} & \textbf{Class C} & \textbf{Class D}\\
		\multicolumn{1}{c|}{} & \textbf{Acuity Matched} & \textbf{Foveally Matched} & \textbf{Peripherally Matched} & \textbf{Non-Acuity Matched} \\
		\hline
		\textbf{Class 1 Fully Foveated} & For any gaze direction, the display meets or exceeds the user’s visual acuity without any peripheral artifacts & For any gaze direction the foveal inset matches user acuity, but peripheral artifacts are present & The foveal inset fails to match user acuity, but achieves equal resolution over all gaze directions with no peripheral artifacts & Neither the foveal inset nor periphery matches user acuity, but the display achieves equal resolution over all gaze directions\\
		\hline
		\textbf{Class 2 Practically Foveated} & For a practical sub-set of gaze directions the display meets or exceeds the user’s visual acuity without any peripheral artifacts & For a practical sub-set of gaze directions the foveal inset matches user acuity $w$ peripheral artifacts present & The foveal inset fails to match user acuity, but achieves equal resolution over a practical sub-set of gaze directions with no peripheral artifacts & Neither the foveal inset nor periphery matches user acuity, but the display achieves equal resolution over a practical sub-set of gaze directions\\
		\hline
		\textbf{Class 3 Partially Foveated} & For a small sub-set of gaze directions the display meets or exceeds the user’s visual acuity without any peripheral artifacts & For a small sub-set of gaze directions the foveal inset matches user acuity $w$ peripheral artifacts present & The foveal inset fails to match user acuity, but achieves equal resolution over a small sub-set of gaze directions with no peripheral artifacts present & Neither the foveal inset nor periphery matches user acuity, but the display achieves equal resolution over a small subset of gaze directions \\
		\hline
		\textbf{Class 4 Non-Foveated} & For a single gaze direction the display meets or exceeds the user’s visual acuity without any peripheral artifacts & For a single gaze direction the foveal inset matches user acuity $w$  peripheral artifacts present & The foveal inset fails to match user acuity and foveal acuity changes with gaze, but no peripheral artifacts are ever present & Neither the foveal inset nor periphery matches user acuity, and the RDF appears to change for any given gaze direction\\
		\hline
	\end{tabular}
\vspace{-1.0em}
	
\end{table*}

Spjut et al. \cite{spjut2020toward} proposed a two-dimensional taxonomy matrix of the foveated display. The first dimension is a resolution-contingent classification, the second is a gaze-contingent classification.
Resolution-contingent classification is based on the acuity distribution function of the human visual model. It describes  howthe non-linear fitting in which the angular resolution perceived by the user decreases as the gaze eccentricity or the angular displacement from the center of gaze increases.

\begin{figure}[htbp]
	\centering
	\includegraphics[width=\linewidth]{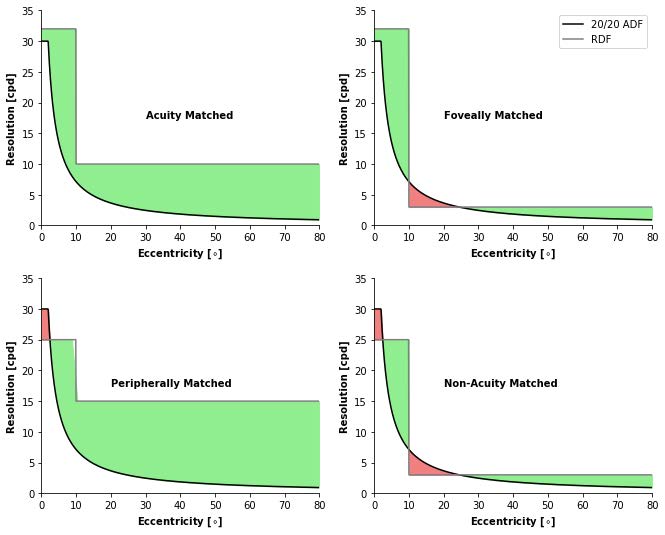}
	\caption{
		{\leftskip=0pt \rightskip=0pt plus 0cm
			Four possible comparisons of the user's visual acuity distribution function and the display's resolution distribution function.
			Images courtesy of Spjut et al. \cite{spjut2020toward}.
		}
	}
	\label{fig:RDF}
\end{figure}

In the resolution classification, the foveated display can be divided into four categories according to the relationship between the visual acuity distribution function (ADF) and the display resolution distribution function (RDF) (Figure \ref{fig:RDF}). 
Class A is acuity matched. This is a conservative display method. The display resolution used in the foveal and peripheral region is higher than the perceptible resolution threshold in the visual acuity distribution function. This type of method ensures the user does not perceive the resolution drop. 
Class B is fovea matched, which means that the display resolution is higher than the user's perceptible threshold in the foveal region. In contrast, the display resolution is lower than the user's perceptible threshold in the peripheral region. To further improve efficiency, this type of method only focuses on the quality of the foveal region. 
Class C is periphery matched, i.e., the user does not perceive any artifacts in the peripheral region, however, the display resolution in the foveal region fails to meet or exceed user visual acuity. 
Class D is non-acuity matched, which means that in the foveal and peripheral region, the display resolution has not reached or exceeded the resolution threshold that can be perceived by human visual acuity. In using this type of method, the user is aware of artifacts in both regions. 
In the second dimension, the gaze-contingent classification, the foveated display can also be divided into four classes according to the gaze direction range in the display. 
Class 1 is the fully foveated display, in which the gaze direction can be any direction in the display. 
Class 2 is the practically foveated display in which the gaze direction should be within $\pm$15$^\circ$ from the center of the display. 
Class 3 is the partially foveated display in which the gaze direction should be much smaller than $\pm$15$^\circ$. 
Class 4 is the non-foveated display in which the single gaze direction is supported.     
Table \ref{tab:2020toward} summarizes the relationship between resolution and gaze-contingent classifications, and provides further detailed descriptions for each.


Weier et al. \cite{weier2017perception} provided an overview of the HVS perceptual mechanisms and classified existing rendering techniques according to different perceptual mechanisms.
Spjut et al. \cite{spjut2020toward} focused on the classification of display effects, which is suitable for hardware display devices and measures the degree of support for foveated rendering by display devices. 

The proposed taxonomy focuses on enabling researchers to easily understand the actual functions, basic ideas, technical framework of current methods and the fundamental design factors that support designers in considering and making technical decisions when designing new methods.
We classify the current foveated rendering methods according to three dimensions: 1) required input data type; 2) foveation principle; 3) rendering paradigm. Table \ref{tbl:vocabulary} shows the elements in each dimension.

\newcommand{\tabincell}[2]{\begin{tabular}{@{}#1@{}}#2\end{tabular}} 
\renewcommand\arraystretch{1.0} 
\begin{table}[htbp]\small
	\setlength{\abovecaptionskip}{5pt}   
	\setlength{\belowcaptionskip}{-5pt}   
	\setlength{\tabcolsep}{4.2mm}{
		\centering
		\caption{
			Taxonomy Vocabulary
		}
		\label{tbl:vocabulary}
		\begin{tabular}{|c|cc|}
			\hline
			\multirow{6}{*}{1. Data Type} & a. & Image/Video \\
			\cline{2-3}   
			& b. & Volume Data \\
			\cline{2-3}  
			& c. & Geometric Meshes \\
			\cline{2-3}  
			& d. & Point Cloud \\
			\cline{2-3}  
			& e. & Hologram Data \\
			\cline{2-3}  
			& f. & Light Field \\
			\hline
			\multirow{6}{*}{\tabincell{c}{2. Foveation\\ Principle}}
			& a. & Multi-spatial Resolution \\
			\cline{2-3}   
			& b. & Multi-temporal Resolution \\
			\cline{2-3}  
			& c. & Multi-luminance Resolution \\
			\cline{2-3}  
			& d. & Multi-color Resolution \\
			\cline{2-3}  
			& e. & Level of Detail \\
			\hline
			\multirow{7}{*}{\tabincell{c}{3. Rendering\\ Paradigm}} 
			& a. & Rasterization \\
			\cline{2-3}   
			& b. & Ray Tracing \\
			\cline{2-3}  
			& c. & Ray Casting \\
			\cline{2-3}  
			& d. & Instant Radiosity \\
			\cline{2-3}  
			& e. & Shadow Mapping \\
			\cline{2-3}  
			& f. & Online/Offline Simplification  \\
			\cline{2-3}  
			& g. &  Neural Rendering  \\
			\cline{2-3}  
			& h. & Photon Mapping\\
			\cline{2-3}  
			& i. & Phase Retrieval\\
			\cline{2-3}  
			& j. & Data Transmission\\
			\hline
		\end{tabular}
	}
\end{table}


Foveated rendering works for different input data. Before understanding or designing a foveated rendering method, it is necessary to consider the processed data type. The input data type is taken as the first dimension of our foveated rendering taxonomy. Present foveated rendering methods can process the data types: image/video, volume data, geometric meshes, point cloud, hologram data, and light field.

The foveation principle is used as the second dimension to classify the previous methods. Foveated rendering provides high-quality rendering for the HVS fovea and provides unnoticeably lower-quality rendering for the periphery. Its core principle is multi-resolution rendering. The present methods use one or several different types of multi-resolution rendering under this ideology, including multi-spatial, multi-temporal, multi-luminance, multi-color, and multi geometry resolution, which is typically referred to as the LoD.

Multi-spatial resolution reduces rendering quality in the output image according to the visual acuity models and the spatial CSFs from the foveal to the peripheral region.
Multi-temporal resolution based methods render one image with multiple resolutions based on spatio-temporal CSFs. Researchers not only consider the HVS spatial error tolerance but also the spatio-temporal error tolerance of dynamic objects and take advantage of the HVS to ensure greater spatio-temporal error tolerance of dynamic objects to effectively perform foveated rendering, which achieves significant improvement in speed \cite{yee2001spatiotemporal, patney2016towards}.
Multi-luminance resolution based methods render one image with multiple resolutions according to spatio-luminance CSFs. Based on the HVS luminance-contrast-awareness, researchers reduce the resolution of peripheral regions with low luminance-contrast sensitivity more aggressively to further improve foveated rendering performance  \cite{tursun2019luminance}.
Foveated rendering alongside the concept of multi-color resolution \cite{duchowski2009spatiochromatic} 
takes advantage of peripheral chromatic degradation, i.e., acceptable peripheral chromatic LoD, and renders one image with multiple color resolutions based on spatio-chromatic CSFs.
Multi-luminance resolution and multi-color resolution based methods are also spatially multi-resolution, however, a particular difference remains in the foveation principle used. To assist readers in more clearly understanding these methods, in this survey, we separated multi-luminance and multi-color resolution based methods from the traditional multi-spatial resolution based methods.
LoD reduces the complexity of the scene geometry in the periphery through visual acuity models and CSFs to reduce computing resources required to render the virtual environment.

The third classification dimension is the rendering paradigm used by existing methods to achieve multi-resolution rendering, which includes: rasterization, ray tracing, ray casting, instant radiosity, shadow mapping, online/offline simplification, photon mapping, neural rendering, and phase retrieval for holographic data. 
For the 360$^{\circ}$ video which is extremely popular in VR applications recently, the encoding, decoding, and transmission mode combined with foveal information directly affect foveated rendering, Thus, we also introduce the data transmission of the 360$^{\circ}$ video as an element of the rendering paradigm.

Table \ref{ltb:summaryOfFoveatedRendering} shows the classification of 90  published reports on the foveated rendering methods from 1990 to 2021 according to the three dimensions.
In addition, the table also lists publications on new devices for foveated rendering  (marked with `-'), the related surveys (*), and the related patent (+).

\section{Early Research from 1990 to 2011} 
\label{sec_earlyWork}

As  foveated rendering research is plentiful spanning a period greater than 30 years, it is divided into two parts organized by chronological order: early research from 1990 to 2011 and recent research over the last 10 years from 2012 to 2021. 

One reason for this is that, with the development of technology, the focus of the recent research has changed compared with early research.

Firstly, the early research in this topic area focused on developing LoD techniques to reduce the complexity of geometric meshes, simulating visual blur effects to enhance
the visual appearance of the rendering results by rasterization on geometric meshes and accelerating the ray casting process for rendering volume data.
With the emergence of Ray Tracing Texel eXtreme (RTX), a high-end professional visual computing platform created by Nvidia that supports real-time ray tracing \cite{alwani2018microsoft}, recent research in foveated rendering paid more attention to accelerating ray tracing for geometric meshes.
Prior to the emergence of RTX, previous ray tracing was only available for non-real-time applications,
such as offline rendering for cinematic visual effects or photo-level realism \cite{sanzharov2020survey}.

\onecolumn


\renewcommand\arraystretch{1.12}
\begin{longtable}[c]{|c|c|c|c|c|c|c|}
	\caption{Summary of Foveated Rendering Technique Implementations (*: survey, +: patent, -: cutting-edge equipment)}
	\label{ltb:summaryOfFoveatedRendering}\\
	\hline
	
	\multicolumn{1}{c|}{Implementation} 
	& \multicolumn{1}{c|}{Data Type} 
	& \multicolumn{1}{c|}{Foveation Principle} 
	& \multicolumn{1}{c}{Rendering Paradigm}   \\
	\hline
	\endfirsthead
	
	\hline
	\multicolumn{1}{c|}{Implementation} 
	& \multicolumn{1}{c|}{Data Type} 
	& \multicolumn{1}{c|}{Foveation Principle}
	& \multicolumn{1}{c}{Rendering Paradigm}  \\
	\hline
	\endhead
	
	\hline
	\endfoot
	
	\hline
	\endlastfoot
	
	\multicolumn{1}{c|}{Levoy 1990 I3D \cite{levoy1990gaze}}
	& \multicolumn{1}{c|}{b} 
	& \multicolumn{1}{c|}{a} 
	& \multicolumn{1}{c}{c}  \\
	\hline
	
	\multicolumn{1}{c|}{Funkhouser 1993 SIGGRAPH \cite{funkhouser1993adaptive}}
	& \multicolumn{1}{c|}{c} 
	& \multicolumn{1}{c|}{e} 
	& \multicolumn{1}{c}{f}  \\
	\hline
	
	\multicolumn{1}{c|}{Ohshima 1996 VR \cite{ohshima1996gaze}}
	& \multicolumn{1}{c|}{c} 
	& \multicolumn{1}{c|}{e} 
	& \multicolumn{1}{c}{f}   \\
	\hline

	
	
	
	\multicolumn{1}{c|}{Luebke 2000 Tech.Rep \cite{luebke2000perceptually}}
	& \multicolumn{1}{c|}{c} 
	& \multicolumn{1}{c|}{a, b, e} 
	& \multicolumn{1}{c}{f}   \\
	\hline
	
	\multicolumn{1}{c|}{Luebke 2001 EG \cite{luebke2001perceptually}}
	& \multicolumn{1}{c|}{c} 
	& \multicolumn{1}{c|}{a, e}
	& \multicolumn{1}{c}{f}   \\
	\hline

	\multicolumn{1}{c|}{Parkhurst 2001 ETRA \cite{parkhurst2001evaluating}}
	& \multicolumn{1}{c|}{c} 
	& \multicolumn{1}{c|}{e} 
	& \multicolumn{1}{c}{f}   \\
	\hline
	
	
	\multicolumn{1}{c|}{Loschky 2001 ARL
		\cite{loschky2001perceptual}}
	& \multicolumn{1}{c|}{a} 
	& \multicolumn{1}{c|}{a} 
	& \multicolumn{1}{c}{a}   \\
	\hline
	
	\multicolumn{1}{c|}{Reddy 2001 CGA
		\cite{reddy2001perceptually}}
	& \multicolumn{1}{c|}{c} 
	& \multicolumn{1}{c|}{b, e} 
	& \multicolumn{1}{c}{f}   \\
	\hline
	
	\multicolumn{1}{c|}{Yee 2001 TOG \cite{yee2001spatiotemporal}}
	& \multicolumn{1}{c|}{c} 
	& \multicolumn{1}{c|}{b} 
	& \multicolumn{1}{c}{a}   \\
	\hline
	
	\multicolumn{1}{c|}{Murphy 2001 EG 
		\cite{murphy2001gaze}}
	& \multicolumn{1}{c|}{c} 
	& \multicolumn{1}{c|}{e} 
	& \multicolumn{1}{c}{f}   \\
	\hline

	\multicolumn{1}{c|}{Parkhurst 2002 HF$^*$ \cite{parkhurst2002variable}}
	& \multicolumn{1}{c|}{/} 
	& \multicolumn{1}{c|}{/} 
	& \multicolumn{1}{c}{/}   \\
	\hline

	\multicolumn{1}{c|}{Cheng 2003 SPA 
		\cite{cheng2003foveated}}
	& \multicolumn{1}{c|}{c} 
	& \multicolumn{1}{c|}{e} 
	& \multicolumn{1}{c}{f}   \\
	\hline
	
	\multicolumn{1}{c|}{Duchowski 2003 Citeseer$^*$ 
		\cite{duchowski2003gaze}}
	& \multicolumn{1}{c|}{/} 
	& \multicolumn{1}{c|}{/} 
	& \multicolumn{1}{c}{/}   \\
	\hline
	
	\multicolumn{1}{c|}{Reingold 2003 SAGE$^*$ \cite{reingold2003gaze}}
	& \multicolumn{1}{c|}{/} 
	& \multicolumn{1}{c|}{/} 
	& \multicolumn{1}{c}{/}   \\
	\hline

	\multicolumn{1}{c|}{Zhou 2004 BDM 
		\cite{zhou2004distance}}
	& \multicolumn{1}{c|}{b} 
	& \multicolumn{1}{c|}{a} 
	& \multicolumn{1}{c}{c}  \\
	\hline

	\multicolumn{1}{c|}{Yu 2005 VC \cite{yu2005fast}}
	& \multicolumn{1}{c|}{b} 
	& \multicolumn{1}{c|}{a} 
	& \multicolumn{1}{c}{c}   \\
	\hline

	\multicolumn{1}{c|}{Lu 2006 EG \cite{lu2006volume}}
	& \multicolumn{1}{c|}{b} 
	& \multicolumn{1}{c|}{a} 
	& \multicolumn{1}{c}{c}   \\
	\hline

	\multicolumn{1}{c|}{Duchowski 2007 TOMM$^*$ 
		\cite{duchowski2007foveated}}
	& \multicolumn{1}{c|}{/} 
	& \multicolumn{1}{c|}{/} 
	& \multicolumn{1}{c}{/}   \\
	\hline

	\multicolumn{1}{c|}{Hillaire 2008 CG\&A \cite{hillaire2008depth}}
	& \multicolumn{1}{c|}{c} 
	& \multicolumn{1}{c|}{a} 
	& \multicolumn{1}{c}{a}   \\
	\hline
	
	\multicolumn{1}{c|}{Hillaire 2008 VR \cite{hillaire2008using}}
	& \multicolumn{1}{c|}{c} 
	& \multicolumn{1}{c|}{a} 
	& \multicolumn{1}{c}{a} \\
	\hline

	\multicolumn{1}{c|}{Duchowski 2009 TAP 
		\cite{duchowski2009spatiochromatic}}
	& \multicolumn{1}{c|}{a}
	& \multicolumn{1}{c|}{d} 
	& \multicolumn{1}{c}{a}  \\
	\hline
	
	\multicolumn{1}{c|}{Murphy 2009 SAP 
		\cite{murphy2009hybrid}}
	& \multicolumn{1}{c|}{c}
	& \multicolumn{1}{c|}{a} 
	& \multicolumn{1}{c}{c}   \\
	\hline
	

	\multicolumn{1}{c|}{Mantiuk 2011 SGDA \cite{mantiuk2011gaze}}
	& \multicolumn{1}{c|}{c} 
	& \multicolumn{1}{c|}{a} 
	& \multicolumn{1}{c}{a}   \\
	\hline

	\multicolumn{1}{c|}{Guenter 2012 TOG \cite{guenter2012foveated}}
	& \multicolumn{1}{c|}{c} 
	& \multicolumn{1}{c|}{a} 
	& \multicolumn{1}{c}{a}   \\
	\hline

	\multicolumn{1}{c|}{Gallo 2013 ISRN \cite{gallo2013high}}
	& \multicolumn{1}{c|}{b} 
	& \multicolumn{1}{c|}{a} 
	& \multicolumn{1}{c}{c}   \\
	\hline
	
	
	\multicolumn{1}{c|}{Duchowski 2014 SAP \cite{duchowski2014reducing}}
	& \multicolumn{1}{c|}{c} 
	& \multicolumn{1}{c|}{a}  
	& \multicolumn{1}{c}{a}  \\
	\hline
	
	\multicolumn{1}{c|}{Fujita 2014 SIGGRAPHAsia \cite{fujita2014foveated}}
	& \multicolumn{1}{c|}{c} 
	& \multicolumn{1}{c|}{a} 
	& \multicolumn{1}{c}{b}   \\
	\hline

	\multicolumn{1}{c|}{Vaidyanathan 2014 Eurographics \cite{vaidyanathan2014coarse}}
	& \multicolumn{1}{c|}{c} 
	& \multicolumn{1}{c|}{a} 
	& \multicolumn{1}{c}{a}   \\
	\hline
	
	\multicolumn{1}{c|}{Mauderer 2014 CHI \cite{mauderer2014depth}}
	& \multicolumn{1}{c|}{c} 
	& \multicolumn{1}{c|}{a} 
	& \multicolumn{1}{c}{a}   \\
	\hline

	\multicolumn{1}{c|}{Patney 2016 SIGGRAPH$^-$ \cite{patney2016perceptually}}
	& \multicolumn{1}{c|}{/}
	& \multicolumn{1}{c|}{/}
	& \multicolumn{1}{c}{/}  \\
	\hline

	\multicolumn{1}{c|}{Patney 2016 TOG \cite{patney2016towards}}
	& \multicolumn{1}{c|}{c} 
	& \multicolumn{1}{c|}{b} 
	& \multicolumn{1}{c}{a}  \\
	\hline
	
	\multicolumn{1}{c|}{Stengel 2016 CGF 
		\cite{stengel2016adaptive}}
	& \multicolumn{1}{c|}{c} 
	& \multicolumn{1}{c|}{b, c} 
	& \multicolumn{1}{c}{a}   \\
	\hline
	
	\multicolumn{1}{c|}{Swafford 2016 SAP \cite{swafford2016user}}
	& \multicolumn{1}{c|}{c} 
	& \multicolumn{1}{c|}{a, e}
	& \multicolumn{1}{c}{a, f}  \\
	\hline
	
	\multicolumn{1}{c|}{Pai 2016 SIGGRAPH$^-$ \cite{pai2016gazesim}}
	& \multicolumn{1}{c|}{/}
	& \multicolumn{1}{c|}{/}
	& \multicolumn{1}{c}{/}  \\
	\hline
	
	\multicolumn{1}{c|}{Lindeberg 2016 
		\cite{lindeberg2016concealing}}
	& \multicolumn{1}{c|}{c} 
	& \multicolumn{1}{c|}{e} 
	& \multicolumn{1}{c}{f}   \\
	\hline
	
	\multicolumn{1}{c|}{MatiasKoskela 2016 ISVC \cite{koskela2016foveated}}
	& \multicolumn{1}{c|}{c} 
	& \multicolumn{1}{c|}{a} 
	& \multicolumn{1}{c}{b}   \\
	\hline

	\multicolumn{1}{c|}{Weier 2016 CGF \cite{weier2016foveated}}
	& \multicolumn{1}{c|}{c} 
	& \multicolumn{1}{c|}{a} 
	& \multicolumn{1}{c}{b}   \\
	\hline

	\multicolumn{1}{c|}{Weier 2017 EG \cite{weier2017perception}}
	& \multicolumn{1}{c|}{/}
	& \multicolumn{1}{c|}{/}
	& \multicolumn{1}{c}{/}  \\
	\hline


	\multicolumn{1}{c|}{Albert 2017 TAP \cite{albert2017latency}}
	& \multicolumn{1}{c|}{a}
	& \multicolumn{1}{c|}{a}
	& \multicolumn{1}{c}{a}  \\
	\hline

	\multicolumn{1}{c|}{Blackmon 2017 USPatent$^+$ 
		\cite{blackmon2017foveated}}
	& \multicolumn{1}{c|}{c} 
	& \multicolumn{1}{c|}{a, e}
	& \multicolumn{1}{c}{a, b, f}  \\
	\hline

	\multicolumn{1}{c|}{Koskela 2017 SIGGRAPH \cite{koskela2017foveated}}
	& \multicolumn{1}{c|}{c}
	& \multicolumn{1}{c|}{a} 
	& \multicolumn{1}{c}{b}   \\
	\hline

	\multicolumn{1}{c|}{Hsu 2017 MM \cite{hsu2017foveated}}
	& \multicolumn{1}{c|}{a} 
	& \multicolumn{1}{c|}{a}
	& \multicolumn{1}{c}{a}  \\
	\hline

	\multicolumn{1}{c|}{Sun 2017 TOG \cite{sun2017perceptually}}
	& \multicolumn{1}{c|}{f} 
	& \multicolumn{1}{c|}{a} 
	& \multicolumn{1}{c}{b}   \\
	\hline

	\multicolumn{1}{c|}{Lungaro 2018 TVCG \cite{lungaro2018gaze}}
	& \multicolumn{1}{c|}{a}
	& \multicolumn{1}{c|}{a} 
	& \multicolumn{1}{c}{j}   \\
	\hline
	
	\multicolumn{1}{c|}{Meng 2018 TOG \cite{meng2018kernel}}
	& \multicolumn{1}{c|}{c} 
	& \multicolumn{1}{c|}{a} 
	& \multicolumn{1}{c}{a}   \\
	\hline
	
	\multicolumn{1}{c|}{MKoskela 2018 CVM \cite{koskela2018instantaneous}}
	& \multicolumn{1}{c|}{c}
	& \multicolumn{1}{c|}{a} 
	& \multicolumn{1}{c}{b}   \\
	\hline

	\multicolumn{1}{c|}{Turner 2018 VR \cite{turner2018phase}}
	& \multicolumn{1}{c|}{c}
	& \multicolumn{1}{c|}{a} 
	& \multicolumn{1}{c}{a}   \\
	\hline
	
	\multicolumn{1}{c|}{Molenaar 2018 \cite{molenaar2018towards}}
	& \multicolumn{1}{c|}{c} 
	& \multicolumn{1}{c|}{a} 
	& \multicolumn{1}{c}{b}   \\
	\hline
	
	\multicolumn{1}{c|}{Weier 2018 TAP \cite{weier2018foveated}}
	& \multicolumn{1}{c|}{c} 
	& \multicolumn{1}{c|}{a} 
	& \multicolumn{1}{c}{b}   \\
	\hline
	
	\multicolumn{1}{c|}{Zheng 2018 VRST \cite{zheng2018perceptual}}
	& \multicolumn{1}{c|}{c} 
	& \multicolumn{1}{c|}{e} 
	& \multicolumn{1}{c}{f}   \\
	\hline

	\multicolumn{1}{c|}{Tan 2018 Opt.Express$^-$ \cite{tan2018foveated}}
	& \multicolumn{1}{c|}{/}
	& \multicolumn{1}{c|}{/} 
	& \multicolumn{1}{c}{/}  \\
	\hline			
	
	\multicolumn{1}{c|}{Wilson 2018 USPatent$^+$ \cite{wilson2018rendering}}
	& \multicolumn{1}{c|}{a} 
	& \multicolumn{1}{c|}{a} 
	& \multicolumn{1}{c}{a}   \\
	\hline

	\multicolumn{1}{c|}{Young 2019 USPatent$^+$ \cite{young2019foveal}}
	& \multicolumn{1}{c|}{d}
	& \multicolumn{1}{c|}{a}
	& \multicolumn{1}{c}{a}  \\
	\hline

	\multicolumn{1}{c|}{Kaplanyan 2019 TOG \cite{kaplanyan2019deepfovea}}
	& \multicolumn{1}{c|}{a} 
	& \multicolumn{1}{c|}{a} 
	& \multicolumn{1}{c}{g}   \\
	\hline

	\multicolumn{1}{c|}{Wei 2019 Appl.Opt. \cite{wei2019fast}}
	& \multicolumn{1}{c|}{e} 
	& \multicolumn{1}{c|}{a} 
	& \multicolumn{1}{c}{b}   \\
	\hline
	
	\multicolumn{1}{c|}{Young 2019 USPatent$^+$ \cite{young2019real}}
	& \multicolumn{1}{c|}{c} 
	& \multicolumn{1}{c|}{e} 
	& \multicolumn{1}{c}{f}   \\
	\hline
	
	\multicolumn{1}{c|}{Tavakoli 2019 USPatent$^+$ \cite{tavakoli2019scene}}
	& \multicolumn{1}{c|}{/} 
	& \multicolumn{1}{c|}{/} 
	& \multicolumn{1}{c}{/}   \\
	\hline
	
	\multicolumn{1}{c|}{Stafford 2019 USPatent$^+$ \cite{stafford2019selective}}
	& \multicolumn{1}{c|}{c} 
	& \multicolumn{1}{c|}{e} 
	& \multicolumn{1}{c}{f}   \\
	\hline
	
	\multicolumn{1}{c|}{Tursun 2019 TOG \cite{tursun2019luminance}}
	& \multicolumn{1}{c|}{c} 
	& \multicolumn{1}{c|}{c} 
	& \multicolumn{1}{c}{b}   \\
	\hline
	
	\multicolumn{1}{c|}{Koskela 2019 EG \cite{koskela2019foveated}}
	& \multicolumn{1}{c|}{c} 
	& \multicolumn{1}{c|}{a} 
	& \multicolumn{1}{c}{b}   \\
	\hline

	\multicolumn{1}{c|}{Ritschel 2019 TOG \cite{friston2019perceptual}}
	& \multicolumn{1}{c|}{c} 
	& \multicolumn{1}{c|}{a} 
	& \multicolumn{1}{c}{a}   \\
	\hline
	
	\multicolumn{1}{c|}{Schutz 2019 VR \cite{schutz2019real}}
	& \multicolumn{1}{c|}{d} 
	& \multicolumn{1}{c|}{e} 
	& \multicolumn{1}{c}{f}   \\
	\hline

	\multicolumn{1}{c|}{Bruder 2019 EuroVis \cite{bruder2019voronoi}}
	& \multicolumn{1}{c|}{b} 
	& \multicolumn{1}{c|}{a} 
	& \multicolumn{1}{c}{c}   \\
	\hline
	
	\multicolumn{1}{c|}{Radkowski 2019 HCII \cite{radkowski2019impact}}
	& \multicolumn{1}{c|}{c} 
	& \multicolumn{1}{c|}{a}
	& \multicolumn{1}{c}{a}  \\
	\hline
	
	\multicolumn{1}{c|}{Siekawa 2019 MMM \cite{siekawa2019foveated}}
	& \multicolumn{1}{c|}{c} 
	& \multicolumn{1}{c|}{a} 
	& \multicolumn{1}{c}{b}   \\
	\hline

	\multicolumn{1}{c|}{Kim 2019 TOG$^-$ \cite{kim2019foveated}}
	& \multicolumn{1}{c|}{/}
	& \multicolumn{1}{c|}{/} 
	& \multicolumn{1}{c}{/}  \\
	\hline
	
	\multicolumn{1}{c|}{Lee 2019 Opt.Express$^-$ \cite{lee2019enhanced}}
	& \multicolumn{1}{c|}{/}
	& \multicolumn{1}{c|}{/} 
	& \multicolumn{1}{c}{/}  \\
	\hline				
	
	\multicolumn{1}{c|}{Bastani 2020 USPatent$^+$ \cite{bastani2020smoothly}}
	& \multicolumn{1}{c|}{c} 
	& \multicolumn{1}{c|}{a} 
	& \multicolumn{1}{c}{a}   \\
	\hline

	\multicolumn{1}{c|}{Spjut 2020TVCG$^*$ \cite{spjut2020toward}}
	& \multicolumn{1}{c|}{/} 
	& \multicolumn{1}{c|}{/} 
	& \multicolumn{1}{c}{/}   \\
	\hline
	
	\multicolumn{1}{c|}{Young 2020 USPatent$^+$ \cite{young2020optimized}}
	& \multicolumn{1}{c|}{c} 
	& \multicolumn{1}{c|}{a} 
	& \multicolumn{1}{c}{a, e}   \\
	\hline
	
	\multicolumn{1}{c|}{Koskela 2020 \cite{koskela2020foveated}}
	& \multicolumn{1}{c|}{c} 
	& \multicolumn{1}{c|}{a} 
	& \multicolumn{1}{c}{b}   \\
	\hline
	
	\multicolumn{1}{c|}{Kang 2020 ACCESS \cite{kang2020depth}}
	& \multicolumn{1}{c|}{b} 
	& \multicolumn{1}{c|}{a} 
	& \multicolumn{1}{c}{c}   \\
	\hline
	
	\multicolumn{1}{c|}{Ananpiriyakul 2020 EI \cite{ananpiriyakul2020gaze}}
	& \multicolumn{1}{c|}{b} 
	& \multicolumn{1}{c|}{a} 
	& \multicolumn{1}{c}{c}   \\
	\hline
	
	\multicolumn{1}{c|}{Wang 2020 ISMAR \cite{wang2020foveated}}
	& \multicolumn{1}{c|}{c} 
	& \multicolumn{1}{c|}{c} 
	& \multicolumn{1}{c}{d}   \\
	\hline

	\multicolumn{1}{c|}{Konrad 2020 TOG \cite{konrad2020gaze}}
	& \multicolumn{1}{c|}{c} 
	& \multicolumn{1}{c|}{a} 
	& \multicolumn{1}{c}{a}   \\
	\hline

	\multicolumn{1}{c|}{Joshi 2020 Access \cite{joshi2020inattentional}}
	& \multicolumn{1}{c|}{c} 
	& \multicolumn{1}{c|}{a} 
	& \multicolumn{1}{c}{a}   \\
	\hline

	\multicolumn{1}{c|}{Meng 2020 TVCG \cite{meng2020eye}}
	& \multicolumn{1}{c|}{c} 
	& \multicolumn{1}{c|}{a} 
	& \multicolumn{1}{c}{a}   \\
	\hline

	\multicolumn{1}{c|}{Meng 2020 TVCG \cite{meng20203d}}
	& \multicolumn{1}{c|}{f} 
	& \multicolumn{1}{c|}{a} 
	& \multicolumn{1}{c}{a}   \\
	\hline
	
	\multicolumn{1}{c|}{Friess 2020 TVCG $^-$\cite{friess2020foveated}}
	& \multicolumn{1}{c|}{/} 
	& \multicolumn{1}{c|}{/} 
	& \multicolumn{1}{c}{/}   \\
	\hline
	
	\multicolumn{1}{c|}{Yoo 2020 OpEx $^-$\cite{yoo2020foveated}}
	& \multicolumn{1}{c|}{/} 
	& \multicolumn{1}{c|}{/} 
	& \multicolumn{1}{c}{/}   \\
	\hline

	\multicolumn{1}{c|}{Bitterli 2020 SIGGRAPH \cite{bitterli2020spatiotemporal}}
	& \multicolumn{1}{c|}{c} 
	& \multicolumn{1}{c|}{b}
	& \multicolumn{1}{c}{c}  \\
	\hline
	
	\multicolumn{1}{c|}{Deza 2021 \cite{deza2021emergent}}
	& \multicolumn{1}{c|}{a} 
	& \multicolumn{1}{c|}{a} 
	& \multicolumn{1}{c}{g}   \\
	\hline
	
	\multicolumn{1}{c|}{Yang 2021 C\&G \cite{yang2021foveated}}
	& \multicolumn{1}{c|}{c} 
	& \multicolumn{1}{c|}{c} 
	& \multicolumn{1}{c}{d}   \\
	\hline

	\multicolumn{1}{c|}{Franke 2021 CGF \cite{franke2021time}}
	& \multicolumn{1}{c|}{c} 
	& \multicolumn{1}{c|}{a,b} 
	& \multicolumn{1}{c}{a}   \\
	\hline

	\multicolumn{1}{c|}{Surace 2021 \cite{surace2021learning}}
	& \multicolumn{1}{c|}{a} 
	& \multicolumn{1}{c|}{a} 
	& \multicolumn{1}{c}{g}   \\
	\hline
	
	\multicolumn{1}{c|}{Youngwook 2021 ISMAR \cite{youngwook2021selective}}
	& \multicolumn{1}{c|}{c} 
	& \multicolumn{1}{c|}{a} 
	& \multicolumn{1}{c}{b}   \\
	\hline
	
	\multicolumn{1}{c|}{Jingyu 2021 ISMAR \cite{jingyu2021perception}}
	& \multicolumn{1}{c|}{c} 
	& \multicolumn{1}{c|}{a} 
	& \multicolumn{1}{c}{b}   \\
	\hline
	
	\multicolumn{1}{c|}{Walton 2021 TOG \cite{walton2021beyond}}
	& \multicolumn{1}{c|}{a} 
	& \multicolumn{1}{c|}{a} 
	& \multicolumn{1}{c}{a}   \\
	\hline
	
	\multicolumn{1}{c|}{Li 2021 TVCG \cite{li2021log}}
	& \multicolumn{1}{c|}{a} 
	& \multicolumn{1}{c|}{a} 
	& \multicolumn{1}{c}{j}   \\
	\hline
	
	\multicolumn{1}{c|}{Shi 2021 TVCG \cite{shi2021foveated}}
	& \multicolumn{1}{c|}{c} 
	& \multicolumn{1}{c|}{c} 
	& \multicolumn{1}{c}{h}   \\
	\hline

	\multicolumn{1}{c|}{Chakravarthula 2021 TVCG \cite{chakravarthula2021gaze}}
	& \multicolumn{1}{c|}{e} 
	& \multicolumn{1}{c|}{a} 
	& \multicolumn{1}{c}{i}   \\
	\hline
	
	\multicolumn{1}{c|}{Jindal 2021 TOG \cite{jindal2021perceptual}}
	& \multicolumn{1}{c|}{c} 
	& \multicolumn{1}{c|}{b,c} 
	& \multicolumn{1}{c}{a}   \\
	\hline

\end{longtable}

\twocolumn


Secondly, recent developments in other technologies have also led to a change in foveated rendering focus. 
For example, cloud rendering became a trend with the development of communication technologies such as 5G, which enables content providers to render 3D programs using a remote server and send back rendered images to user terminals interactively \cite{mukhina2015method}. 
Cloud rendering has revolutionized foveated rendering-based data transmission. The development of deep learning techniques has also used foveated images or videos to improve the accuracy of deep learning models for specific computer vision tasks. These key developments have initiated vital research hotspots in the field.

Thirdly, researchers proposed new data types, such as point cloud, hologram data, and light fields, to meet the requirements of different applications. 
Incorporating the rendering paradigm for these new data types into the foveated rendering framework has also become a critical research area.

Another reason for this division is that readers may have different requirements for early and recent research. For the former, typically readers solely require understanding of the methods function and fundamental ideas. While for the latter, because it is the state of the art, it may be necessary to reproduce and compare recent research, such that readers can establish a deeper understanding of contemporary foveated rendering. 

Foveated 3D graphics \cite{guenter2012foveated} proposed in 2012 is an essential milestone for dividing research on the topic into two parts. This introduced a rasterization-based foveated rendering system to improve rasterization rendering performance, demonstrating that users cannot perceive the degradation of rendering quality from foveated rendering in this system because of the publication of extremely detailed perceptual experiments. Prior to this, foveated rendering primarily mimicked HVS visual effects to improve the visual appearance of images. In subsequent research, foveated rendering focused on improving rendering performance without perceptual loss.

\begin{figure}[htbp]
	\centering
	\includegraphics[width=\linewidth]{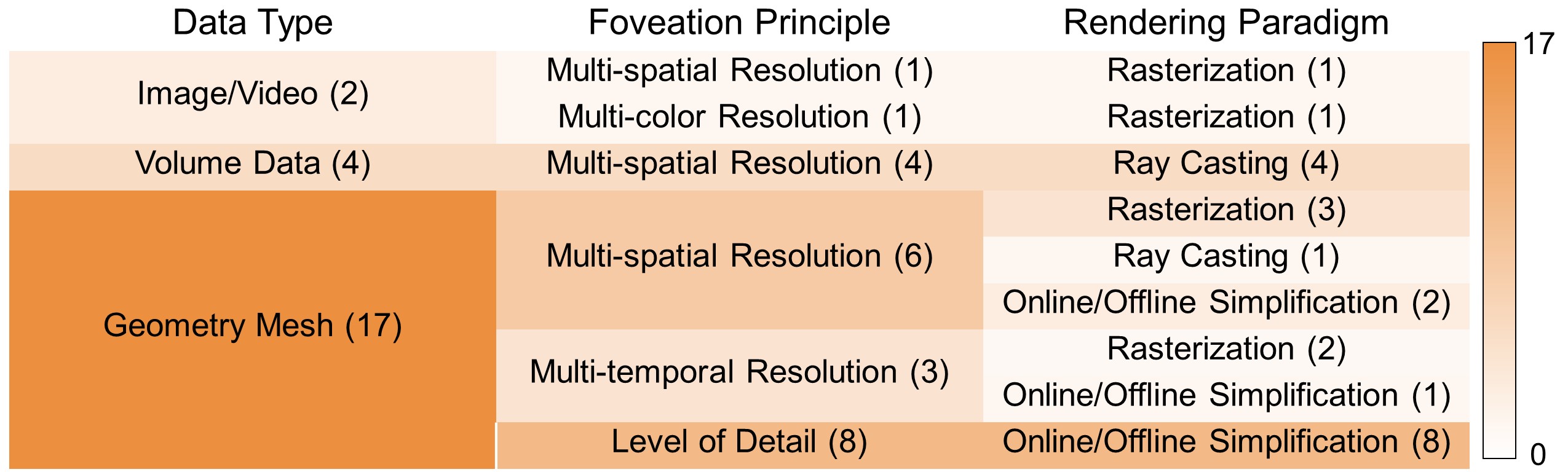}
	\caption{
		{\leftskip=0pt \rightskip=0pt plus 0cm
			Frequency of research in foveated rendering from 1990 to 2011.
			Numbers in parentheses indicate the number of studies using the specific data type, foveated principle and rendering paradigm are listed in front of parentheses.
		}
	}
	\label{fig:statics_papers_early}
\end{figure}

In this section, early foveated rendering research is reviewed. Figure \ref{fig:statics_papers_early} visualizes the frequency of various research on the topic from 1990-2011 according to the proposed taxonomy.
We initially summarized discussions in review papers from 1990 to 2011 (Section \ref{sec_early_review}).
Subsequently, we introduced the methods according to their frequency of occurrence from high to low: 
1) foveated rendering based on LoD  (Section \ref{sec_early_lod}); 
2) foveated rendering based on multi-spatial resolution for volume data (Section \ref{sec_early_msrVol}); 
3) foveated rendering based on multi-spatial resolution for geometric meshes (Section \ref{sec_early_msrGeo}).
Additionally, some techniques closely related to foveated rendering in early research are introduced(Section \ref{sec_early_otherRW}).
In early research, foveated rendering was also referred to as gaze-directed rendering (GDR), gaze-contingent rendering (GCR), or gaze-contingent display (GCD).  

\subsection{Reviews}
\label{sec_early_review}

Several reviews discuss and summarize the early foveated rendering research. 

For example, Reingold et al. \cite{reingold2003gaze} discussed gaze-contingent multi-resolution displays (GCMRD)  indifferent areas, including engineering design research on the development of GCMRDs, multi-resolution image processing, multi-resolution sensors, human factors research on multi-resolution displays, gaze-contingent displays, and human-computer interaction. Focus was placed on reviewing methods to solve two questions regarding gaze-contingent multi-resolution displays: 1) image degradation owing to the characteristics of multi-resolution images, vision model based multi-resolution images generation methods, discrete/continuous-resolution drop-off, and color resolution drop-off were reviewed; 2) for perceptible image motion caused by image updating, gaze/head/hand-contingent displayed area of interest (D-AOI) movement-based methods and predictive D-AOI movement-based methods were analyzed. Parkhurst et al. \cite{parkhurst2002variable} reviewed variable-resolution displays from the three aspects: 1)  potential computational savings achieved with variable-resolution displays; 2) practical constraints in implementing variable-resolution displays; 3) the behavioral consequences of using variable-resolution displays, such as perceptual quality, task performance, and eye movement measures. 
The authors also explained that gaze-related rendering in virtual reality is only one variable-resolution display application. Variable-resolution displays could also be used in low-vision enhancement and internet image transmission applications. Duchowski et al.\cite{duchowski2003gaze} divided gaze-contingent display methods into two categories: model-based graphical displays and screen-based displays. Model-based methods used the objects' LoD to generate the image matching the resolvability of the human retina, while the screen-based methods adjusted the image quality at the pixel level. Focus plus context methods were also discussed, which were extremely similar to foveal and peripheral displays. Duchowski et al. \cite{duchowski2007foveated} reviewed perceptually loss-less gaze-contingent displays, space-variant imaging based on the pyramidal idea, and gaze-contingent displays for stereoscopic imaging. The authors also summarized GPU-based gaze-contingent displays before 2007 and introduced related technologies including mipmapping, multitexturing and fragment programming.

\subsection{LoD}
\label{sec_early_lod}

Clark et al. \cite{clark1976hierarchical} introduced the concept of discrete LoD, which defined several versions of the model at different levels, using a detailed grid when the object is close to the observer and replacing it with a coarser approximation when the object is is far from the observer. 
The LoD technique can be combined with foveated rendering to reduce the complexity of scenes according to the user's gaze position and the perceptual models, which significantly improves time performance \cite{luebke2003level}.
Funkhouser et al. \cite{funkhouser1993adaptive} proposed a gaze-directed dynamic LoD selection system that considers motion blur and visual acuity.
The motion blur value is expressed by the speed at which the object image moves on the retina.
The visual acuity value is expressed by the distance from the object to the center of the user's gaze.
Owing to the lack of an accurate perceptual model, the effect of motion blur is controlled by a slider set by the user. 
As there is no eye-tracking system, the user's gaze is assumed to be at the center of the screen. 
This research firstly introduced the concept of gaze-directed perceptual LoD.
Ohshima et al. \cite{ohshima1996gaze} used the ultrasonic sensors built into the eye-trackers to measure head direction, which is used as a substitute for gaze direction.
The authors introduced a visual acuity fall-off model, a binocular horopter model, and a kinetic vision model, respectively, to calculate visual acuity according to eye direction, and subsequently mapped the minimum visual acuity calculated by the three models to control the LoD for rendering.

As the discrete LoD technique cannot locally change details, for example, the side of a large object near the view cannot be rendered in great detail while simultaneously reducing its distant details. Rather than calculating a series of static LoDs in the pre-process, Hoppe et al. \cite{hoppe1996progressive} introduced the concept of continuous LoD. They built a data structure from which the desired LoD can be extracted at runtime.
In foveated rendering,
Luebke et al. \cite{luebke2000perceptually} proposed a gaze-directed continuous LoD framework. They employed a commercial eye tracker to measure the user's gaze over a desktop display in real time, then introduced a perceptual metric to measure the level of geometric meshes based on the visual acuity fall-off model proposed in \cite{rovamo1979estimation} and the spatio-temporal contrast sensitivity function proposed in \cite{kelly1979motion}.
Murphy et al. \cite{murphy2001gaze} employed a binocular eye-tracked VR system to obtain the gaze in VR, then modeled visual acuity fall-off for both eyes based on the gaze. Subsequently, they proposed the gaze-contingent continuous LoD to degrade the resolution of meshes based on visual acuity.
Luebke et al. \cite{luebke2001perceptually} provided a perception based node fold system for the vertex tree, which is a hierarchical clustering of vertices. They identified that the perceptible result of a change induced by simplification can be conservatively equal to the change of its lowest spatial frequency and maximum contrast. Thus the perception-based node expansion system visits each node in the vertex tree top-down. If the lowest spatial frequency and maximum contrast induced by folding the node are less than the pre-defined threshold contrast, the system folds the node. Otherwise, the node will remain unfolded, and traversal continues.
Reddy et al. \cite{reddy2001perceptually} noted that previous perceptually based LoD research used pre-simplified versions of an object that can be selected for rendering in a view-dependent manner. They performed a per-pixel calculation of the pixel's spatial frequency by employing the GPU, then used the spatial frequency to determine the LoD based on an eccentricity-based spatio-temporal CSF \cite{kelly1979motion}.
Parkhurst et al. \cite{parkhurst2001evaluating} conducted virtual search tasks to evaluate straight-forward gaze-contingent continuous LoD rendering, in which the LoD decreases linearly as the distance from the rendered object to the point of gaze increases. The results demonstrated that the behavioral performance gains could offset behavioral performance costs of gaze-contingent LoD techniques owing to increased rendering performance. 
Cheng et al.\cite{cheng2003foveated} used surface information obtained from a 3D scanner and allowed a user to select a foveal point, then proposed an interactive LoD update with foveation.

\subsection{Multi-spatial Resolution for Volume Data}
\label{sec_early_msrVol}
Rendering volume data inherently consumes massive computing resources owing to large data size. Thus real-time rendering of large volume datasets was infeasible using desktop personal computers in earlier years. One solution is to use ray casting to render volume data based on the concept of multi-spatial resolution, i.e., to render objects in the foveal region at full resolution and ignore details of objects in the peripheral region, which can reduce calculation and communication requirements.

\begin{figure}[htbp] 
	\centering
	\includegraphics[width=\linewidth]{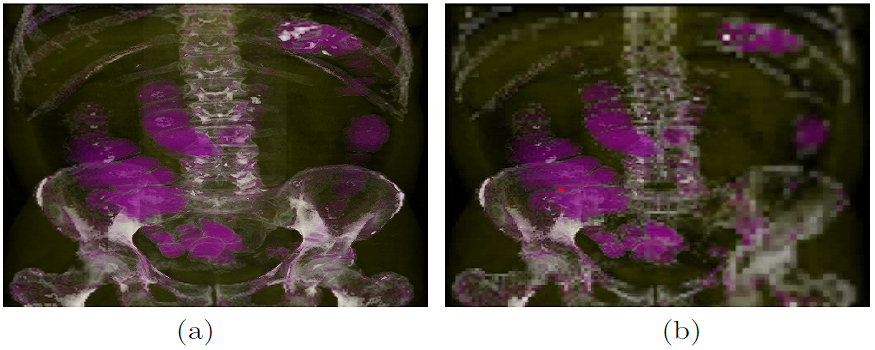}
	\caption{
		{\leftskip=0pt \rightskip=0pt plus 0cm
			Fast rendering of foveated volumes in wavelet-based representation proposed by Yu et al. \cite{yu2005fast}. (a) is rendered with a full resolution, (b) is rendered with this method, the fovea is situated at the red dot. This method achieved a 1.3-8$\times$ improvement in speed compared with the full resolution ray casting.
			Images courtesy of Yu et al. \cite{yu2005fast}.
		}
	}
	\label{fig:rw_2005fast}
\end{figure}

Levoy et al. \cite{levoy1990gaze} first explored the method for incorporating foveated rendering into volume rendering. They used the Eye-Mark eye tracker to obtain the user's gaze direction and directed this at an object by rotating the user's eyes or head until the object's projection falls on the fovea.
Subsequently, they distributed the number of casting rays per unit area and the number of samples taken along the unit length of each ray based on a visual acuity fall-off model. 
For weakening unnecessary objects in the peripheral region, Zhou et al. \cite{zhou2004distance} adjusted the opacity of the sample according to the distance from the sample point to the center of the foveal region for volume feature enhancement, which assisted users in focusing more on objects in the foveal region. 
To further accelerate foveated volume rendering, Yu et al. \cite{yu2005fast} remapped the mask which was used to sample the rays and the length of each ray into a small number of wavelet coefficients in the wavelet domain according to the visual acuity fall-off model. 
Figure \ref{fig:rw_2005fast} visualizes the rendering results of full-resolution ray casting and the proposed method.
Lu et al. \cite{lu2006volume} used a camera to focus on one eye and record eye movements as the user observes the volume, and employed the eccentricity-based spatio-temporal CSF \cite{kelly1979motion} to acquit the HVS importance information,  subsequently, they used this importance information to fix object shapes, positions and to tune opacity transfer functions automatically.

\subsection{Multi-spatial Resolution for Geometric Meshes} \label{sec_early_msrGeo}




In addition to volume data, the concept of multiple spatial resolution is also used to accelerate geometric mesh rendering.

Murphy et al. \cite{murphy2009hybrid} proposed a hybrid technique based on the visual acuity fall-off model and the spatial CSF, which used ray casting to sample the scene's geometry. This technique enables non-isotropic degradation within meshes without directly manipulating mesh geometry.

As early geometry models were coarse, geometric mesh performance rendering in the entire image at high resolution is acceptable. 
Thus, researchers in foveated rendering focused more on simulating the HVS visual appearance, i.e., gaze-contingent depth-of-field (DoF) rendering, rather than accelerating geometric mesh rendering.
The traditional pinhole camera model in computer graphics can sharply present objects at all distances. However, in the eyes and real cameras, only objects within the focal range can be sharply displayed, while objects far away or close to the viewpoint are blurred. To simulate the fact that humans only perceive sharp objects within a certain distance range near the focal length and to improve the user's immersion, gaze-contingent DoF rendering was introduced in Hillaire et al.  \cite{hillaire2008using} and Mantiuk et al. \cite{mantiuk2011gaze}. 
In this review, we regard gaze-contingent DoF as a specific type of foveated rendering, which pays more attention to the scene depth range in the foveal region.

For improving gaze-contingent DoF perception during first-person navigation in virtual environments (VE), 
Hillaire et al. \cite{hillaire2008depth} proposed a gaze-contingent DoF blur filter which simulates the blurring of objects located in front of or behind the focus point of the eyes, and a peripheral blur filter which simulates the blurring of objects situated in the periphery of the field of vision.
Hillaire et al. \cite{hillaire2008using} subsequently described an algorithm for calculating the focal length and point in the 3D virtual environment and used the gaze-contingent DoF blur and peripheral blur filters proposed in Hillaire et al. \cite{hillaire2008depth} to render the DoF blur effects to simulate the fact that humans only perceive sharp objects within a certain distance range near the focal length. Mantiuk et al. \cite{mantiuk2011gaze} evaluated human impression regarding the existence of the DoF phenomenon in the 3D virtual environment. The results demonstrated that people noticed and preferred the DoF visualization controlled by the eye tracker. The best impression was achieved with the medium blurriness level (the lens aperture diameter was 7$cm$).

In early foveated rendering research, researchers also adopted concepts of multi-color and multi-temporal resolution in foveated rendering. Duchowski et al. \cite{duchowski2009spatiochromatic} investigated the color reduction in the peripheral region. The results demonstrated that peripheral chromaticity could not be reduced within the central 20$^\circ$ visual angle, i.e., the color reduction should be maintained isotropically across the central 20$^\circ$ visual field. 

\subsection{Other Related Work}
\label{sec_early_otherRW}

From 1990 to 2011, some other related foveated rendering research emerged, such as \textit{perception based rendering}, \textit{focus+context visualization}, \textit{selective rendering}, and \textit{multi-resolution display}.






\textit{Perception-based rendering} refers to use of the HVS features and associated perceptual models to improve rendering performance and to enhance the perceptual quality of rendering results.
For example, Yee et al. \cite{yee2001spatiotemporal} constructed a spatio-temporal error tolerance map based on a spatio-temporal CSF that accepts low-quality rendering in highly error-tolerant regions without degrading perceptual quality, thus improving rendering speed.
Unlike perception-based rendering, all HVS features and perceptual models in foveated rendering are highly related to the HVS foveal features.

In early research involving perception-based rendering, researchers conducted user studies of perceptual models to obtain useful parameters and error metrics that have a direct impact on foveated rendering.
For example, Ramasubramanian et al. \cite{ramasubramanian1999perceptually} introduced an error metric considering the spatial-luminance CSF, which predicted the perceptual threshold to detect artifacts in 3D scenes. Karol Myszkowski et al. \cite{myszkowski2001perception} presented a perceptual error metric based on a spatio-temporal CSF, which retained inherent noise in the animation generated using stochastic methods below human observer sensitivity.

\textit{Focus+context visualization} is a rendering technique that visualizes more critical information by removing or suppressing less critical parts of the scene. Critical information typically has semantic integrity.
Focus+context visualization typically uses distortion and highlighting to visualize interested objects in focus and nearby related objects in context \cite{furnas1986generalized, sarkar1992graphical, lamping1995focus, carpendale1996distortion, plaisant2002spacetree, kosara2002focus, munzner2003treejuxtaposer, viola2004importance}, 
while foveated rendering is based on HVS perception theories to allocate further computing sources to the foveal region.

Carpendale et al. \cite{carpendale1996distortion} highlighted data by dedicating additional space to this and applied distortions to abstract graphs to observe interested graph nodes clearly. 
Viola et al. \cite{viola2004importance} proposed a view-dependent model for automatic focus+context volume visualization. This model enables interested objects to be displayed more accurately to view further details, while occluded objects are displayed with low accuracy or completely suppressed.

\textit{Selective rendering} is task-dependent rendering, which uses HVS knowledge to select the objects in scenes that require rendering based on application tasks \cite{cater2002selective, cater2003detail, sundstedt2004selective, sundstedt2004top}, i.e., different tasks require different objects to be drawn. For example, if the task is to count the number of pencils in a mug on a table in a room, only the image in the visual angle of the fovea centered around the pencils is rendered with high quality. Cater et al. \cite{cater2002selective} designed perceptual experiments to prove that users would ignore parts of the scene that were not related to a specific task, which can be used to reduce rendering time without affecting visual quality in interactive tasks.
Sundstedt et al. \cite{sundstedt2004selective, sundstedt2004top} investigated the extent to which image resolution, edge anti-aliasing and reflection, and shadow parameters can be reduced between non-task-related and task-related regions when viewers cannot perceive image quality degradation.

\textit{Multi-resolution Display} focused on a more general pipeline of multi-resolution rendering \cite{duchowski1995simple, geisler1998real, geisler1999variable, parkhurst2000evaluating, geisler2002real}. 
In addition to foveated rendering, the multi-resolution display can also be used for perception-based and selective rendering etc.
Duchowski et al. \cite{duchowski1995simple} introduced a multi-resolution display method based on mipmap texture mapping. They retained the original image resolution in multiple regions of interest (ROIs) selected by users and gradually reduced the periphery around each ROI according to the specified resolution mapping function.
Geisler et al. \cite{geisler1998real} developed a foveated multi-resolution pyramid video coding/decoding system that uses a foveated multi-resolution pyramid to encode each image into five or six regions of different resolutions and eliminated spatial edge artifacts between the regions generated by foveation through raised-cosine blending across levels of the pyramid and “foveation point interpolation” within pyramid levels.
Geisler et al.\cite {geisler1999variable} described a multi-resolution pyramid method that used a pyramid encoder to divide the image into 2-6 layers, and used a pyramid decoder to sample each layer at different rates.
Parkhurst et al. \cite{parkhurst2000evaluating} introduced a two-region gaze-contingent display and investigated behavioral effects on the display based on a visual search task.
They identified that reaction time and accuracy co-vary as a function of the foveal region size.
For the small foveal region, slow reaction times are accompanied by high accuracy. Conversely, for the large foveal region, fast reaction times are accompanied by low accuracy.
Geisler et al. \cite{geisler2002real} proposed a method to generate completely arbitrary variable-resolution displays based on image pyramidal pre-processing\cite{geisler1998real}.

	\section{Foveated Rendering over the Past Decade (2012-2021)} \label{sec_recentWork}

\begin{figure}[htbp]
	\centering
	\includegraphics[width=\linewidth]{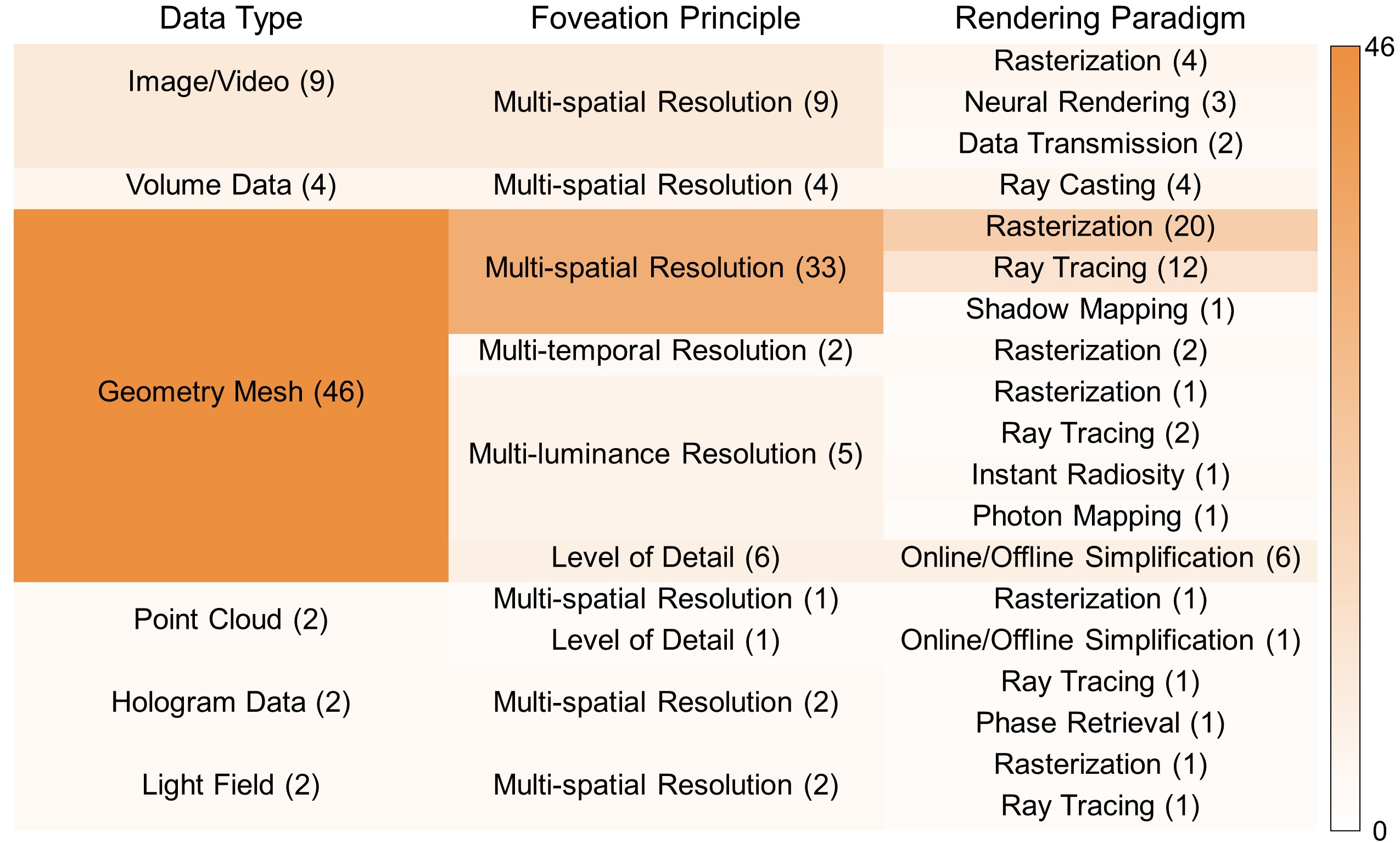}
	\caption{
		{\leftskip=0pt \rightskip=0pt plus 0cm
			Frequency of research in foveated rendering from 2012 to 2021.
			Numbers in parentheses indicate the number of studies using the specific data type, foveated principle and rendering paradigm are listed in front of parentheses.
		}
	}
	\label{fig:statics_papers_recent}
\end{figure}

This section reviews foveated rendering research published most recently over the past decade. Figure \ref{fig:statics_papers_recent} visualizes the frequency of various foveated rendering research from 2012 to 2021 according to the proposed classification method. LoD or multi-spatial resolution rasterization methods for geometric meshes, and multi-spatial resolution methods for volume data remain research hotspots. Furthermore, methods such as multi-spatial resolution ray tracing for geometric meshes, and multi-spatial resolution methods for images or videos have also attracted keen attention from researchers. In the following subsections, we introduce methods in these classes according to the frequency of occurrence from high to low: 
1) foveated rendering based on multi-spatial resolution, rendering geometric meshes with rasterization (Section \ref{sec_recent_MSR_mesh_ras}); 
2) foveated rendering based on multi-spatial resolution, rendering geometric meshes with ray tracing (Section \ref{sec_recent_MSR_mesh_rt}); 
3) foveated rendering based on multi-spatial resolution, rendering image/video data (Section \ref{sec_recent_MSR_img}); 
4) foveated rendering based on LoD (Section  \ref{sec_recent_lod}); 
5) multi-spatial resolution for volume data (Section \ref{sec:vol});
6) multi-luminance resolution method for geometric meshes (Section \ref{sec: mol});
and 7) foveated rendering for nascent data types  (Section \ref{sec_recent_MSR_other}).


\subsection{Multi-spatial Resolution Rasterization for Geometric Meshes}
\label{sec_recent_MSR_mesh_ras}

In recent years, with the development of modeling technology, the complexity of 3D models and the scale of virtual scenes have increased. In multiple virtual reality applications, using high-resolution and high-quality rasterization of the scene cannot achieve real-time frame rates. Therefore, many researchers focused on the foveated rendering method alongside improving geometric mesh rasterization performance based on the concept of multi-spatial resolution.

\begin{figure}[htbp] 
	\centering
	\includegraphics[width=\linewidth]{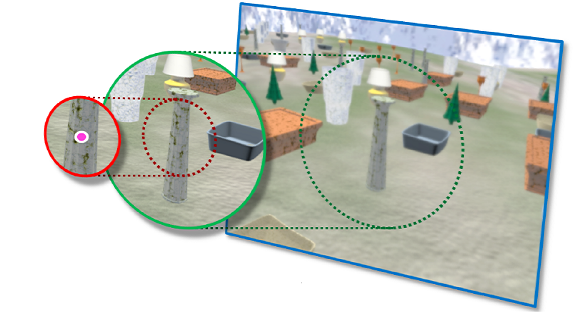}
	\caption{
		{\leftskip=0pt \rightskip=0pt plus 0cm
			Foveated 3D graphics proposed by Guenter et al. \cite{guenter2012foveated}. Three nested layers were rendered (red, green, and blue) at three different resolutions through rasterization based on a visual acuity fall-off model. The three nested layers are combined to generate the final image. This method could achieve comparable perceptual quality with reference to  traditional full-resolution rendering, but at a 4-6.2$\times$ speed improvement.
			Images courtesy of Guenter et al. \cite{guenter2012foveated}.
		}
	}
	\label{fig:rw_2012foveated}
\end{figure}

Guenter et al. \cite{guenter2012foveated} took advantage of the visual acuity fall-off model and rendered three nested layers by rasterization. The pipeline for this method is described in Figure \ref{fig:rw_2012foveated}. These nested layers are rasterized as the angular diameter decreases in resolution to achieve improved rendering performance. Finally, three layers are mixed to form the final image. The results demonstrate that the rendering speed of this method is 5-6$\times$ that of the traditional method. The quality users visually perceive is comparable to traditional rendering.
Vaidyanathan et al. \cite{vaidyanathan2014coarse} presented a novel architecture to flexibly control shading rates in a rasterization pipeline named Coarse Pixel Shading (CPS) and tested the architecture for foveated rendering with a visual acuity fall-off model. 
As CPS pipelines require adaptive shading features not yet commonly available on commodity GPUs, Meng et al. \cite{meng2018kernel} presented a simple two-pass kernel foveated rendering (KFR) pipeline that maps well onto modern GPUs. In the first pass, they computed the kernel log-polar transformation and rendered it to a reduced-resolution buffer. The second pass carried out the inverse-log-polar transformation with anti-aliasing to map reduced-resolution rendering to the full-resolution screen. The results showed that KFR could achieve a 2.8-3.2$\times$ speed improvement in rendering on 4K UHD (2160$p$) displays with less perceptual LoD. 

\begin{figure*}[htbp]  
	\centering
	\includegraphics[width=\linewidth]{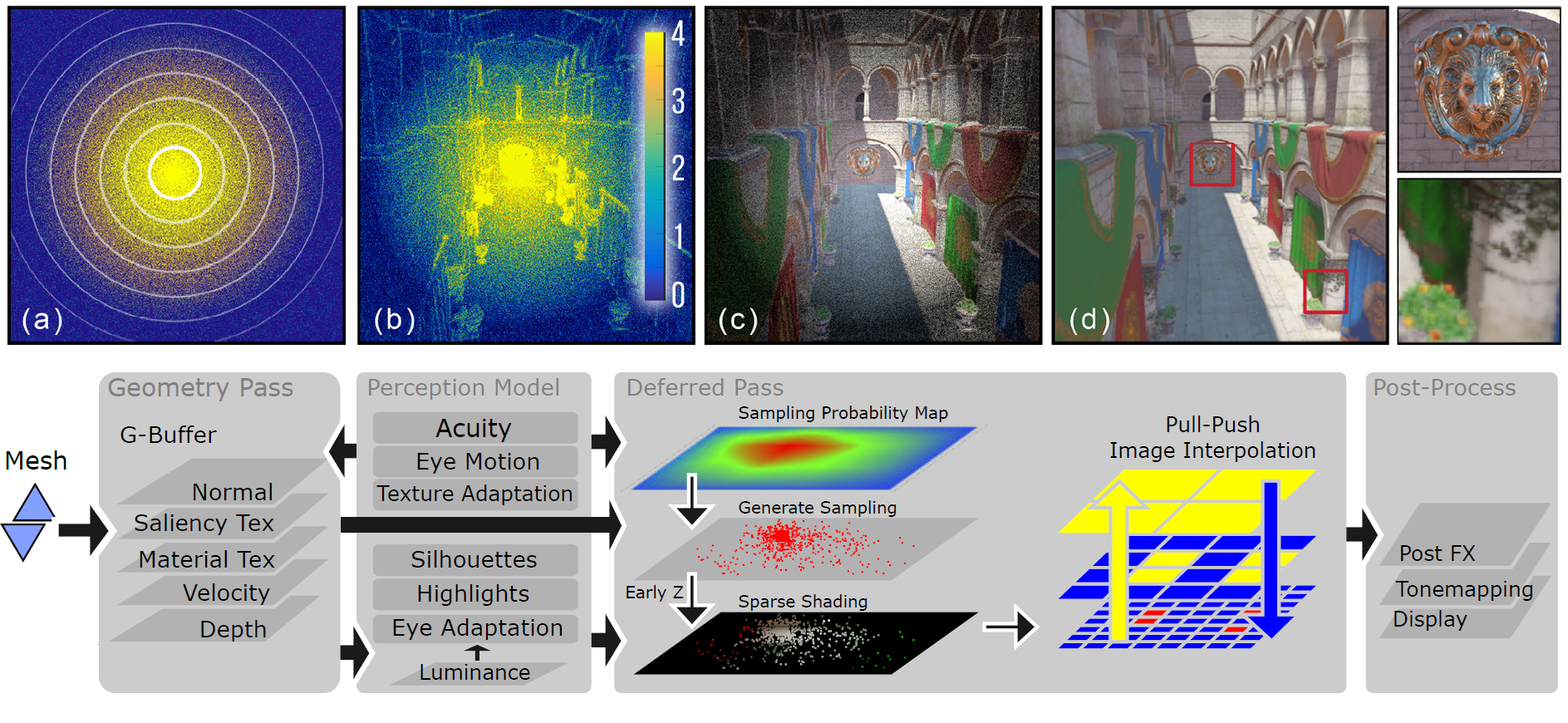}
	\caption{
		{\leftskip=0pt \rightskip=0pt plus 0cm
			Adaptive image-space sampling method for foveated rendering proposed by Stengel et al. \cite{stengel2016adaptive}. 
			A perceptual adaptive sampling pattern (b) was constructed for sparse shading (c), which combined visual cues such as visual acuity (a), spatial, spatio-temporal, and spatio-luminance CSFs. Fast image interpolation was performed in the periphery (d) to achieve the same perceptual quality with less shading cost.
			Row 2 shows the pipeline of the proposed method: in the geometry pass, this generates the G-Buffer; in the deferred pass, it first generates the sampling pattern, then performs sparse shading based on the sampling pattern, and finally uses a pull-push operation to complete the missing image parts by interpolation; in the post-processing pass, it applies post-processing operations similarly to tone mapping and grading before displaying the final image.
			The final image contains high details in the fovea and low details in the periphery.
			Images courtesy of Stengel et al. \cite{stengel2016adaptive}.
		}
	}
	\label{fig:rw_2016adaptive}
\end{figure*}

In addition to considering the spatial factor, much research considered the temporal factor, based on the concept of multi-temporal resolution to further accelerate geometric mesh rasterization.
Stengel et al. \cite{stengel2016adaptive} introduced a sampling method based on the visual acuity fall-off model, the spatio-temporal and the spatio-luminance CSFs, and subsequently integrated the sampling method into the deferred shading pipeline. Only important image features were shaded while interpolating the remaining features without affecting perceived quality. The visualization results are shown in Figure \ref{fig:rw_2016adaptive}. 
Patney et al. \cite{patney2016towards} designed a foveated rendering system that reduces the number of shadings by up to 70\%, the authors subsequently introduced a novel anti-aliasing algorithm based on a visual acuity fall-off model and a spatio-temporal CSF. This anti-aliasing algorithm assists in recovering peripheral region details that are resolvable by human eyes albeit degraded by filtering.
Franke et al. \cite{franke2021time} presented a foveated rendering method that comprised recycling pixels in the periphery by spatio-temporally reprojecting them from previous frames to accelerate rendering performance. This reprojection detected and re-evaluated artifacts and disocclusions according to a confidence value determined by a perception-based metric.
Jindal et al. \cite{jindal2021perceptual} proposed the variable-rate shading pipeline to accelerate rasterization rendering performance 
This approach divides the output image into a number of 16$\times$16 image tiles, 
and subsequently adaptively adjusts the shading accuracy and refresh rate of each image tile based on spatio-temporal and the spatio-luminance CSFs.

To further improve calculation process speed, Turner et al. \cite{turner2018phase} aligned the rendered pixel grid to virtual scene content during rasterization and upsampling, which reduced the detectability of motion artifacts in the periphery without complex interpolation or anti-aliasing algorithms.
Bastani et al. \cite{bastani2020smoothly} rendered an intermediary image of the 3D scene in the intermediary compressed space and unwarped the image to generate the foveated image.
Young et al.\cite{young2020optimized} adopted foveated rendering to accelerate shadow rendering. Shadow mapping was used to obtain two shadow maps of different resolutions and geometric meshes in the foveal region were rendered with the high-resolution shadow map, while that of the peripheral region were rendered using the low-resolution shadow map.

Towards HMDs with latency and field-of-view requirements, Friston et al. \cite{friston2019perceptual} presented a rasterization pipeline that achieved foveated rendering in one rasterization pass with per-fragment ray-casting. 
Meng et al. \cite{meng2020eye} accelerated foveated rendering on HMDs with more aggressive foveation based on the theory of ocular dominance.

Foveated rendering improves the frame rate and quality of foveal vision by reducing peripheral vision resolution. However, foveated rendering optimization is a difficult task. This requires careful selection of multiple parameters, such as the number of layers, eccentricity, resolution of the peripheral region, and foveated rendering perceptibility must be evaluated. Therefore, many researchers designed perceptual studies to optimize and evaluate the task.
Patney et al. \cite{patney2016towards} designed a user study to evaluate users’ perceptual abilities of peripheral vision when viewing today’s displays. The results demonstrated: 1) filtering peripheral regions would reduce contrast, thereby creating a sense of tunnel vision; 2) when applying the post-processing contrast enhancement function, the object could tolerate a 2$\times$ larger blur radius before detecting the difference from the non-foveated ground truth.
Swafford et al. \cite{swafford2016user} applied foveated rendering to the multi-resolution, screen-space ambient occlusion, and tessellation methods. Practical rules for each method were proposed to achieve significant performance gains with user studies and the newly proposed rendering quality metrics.


Recent research also concentrated on gaze-contingent DoF rendering based on the concept of multi-spatial resolution. Mauderer et al. \cite{mauderer2014depth} designed a user study to demonstrate that gaze-contingent DoF increased subjective perceived realism and depth and could contribute to the perception of ordinal depth and distance between objects, however, it was limited in accuracy.
Duchowski et al. \cite{duchowski2014reducing} used gaze-contingent DoF to reduce users' visual discomfort when viewing stereoscopic displays.  However, similar to earlier attempts, participants disliked gaze-contingent DoF, which may be attributed to eye tracker spatial inaccuracy and the DoF simulation's noticeable temporal lag.
Konrad et al. \cite{konrad2020gaze} extended gaze-contingent DoF rendering to ocular parallax rendering, which described the small amounts of depth-dependent image shifts on the retina created as the eye rotates. They introduced ocular parallax rendering technology that accurately rendered small amounts of gaze-contingent parallax capable of improving depth perception and realism in VR. The results demonstrated that ocular parallax rendering provided an effective ordinal depth cue and improved the impression of realistic depth in VR.
Walton et al. \cite{walton2021beyond} believed that the HVS perceives the that periphery is more than just blurry, and proposed a real-time method to compute images identical to ground truth images in terms of peripheral perception.

In addition, researchers applied foveated rendering to VR interaction. 
Joshi et al. \cite{joshi2020inattentional} presented foveated rendering-based redirected walking in VR, which capitalized on naturally occurring saccades and blinks to completely refresh the framebuffer.
Radkowski et al. \cite{radkowski2019impact} conducted a user study to demonstrate whether the foveated rendering technique would distract users and reduce their training effect in VE. The results demonstrated that the user noticed the technology but was not negatively affected by it, and the performance difference was insignificant, except for some outliers caused by technical eye-tracking limitations.

In addition to geometric meshes, multi-spatial resolution rasterization can also be used for foveated rendering on point clouds \cite{young2019foveal}.



\subsection{Multi-spatial Resolution based Ray Tracing}
\label{sec_recent_MSR_mesh_rt}
Ray tracing is capable of controlling the number of rays emitted from each pixel. The more rays emitted from a single pixel, the higher the rendering quality of that pixel. Therefore, the ray tracing framework naturally supports spatial multi-resolution rendering. Koskela et al. \cite{koskela2016foveated} provided a theoretical estimation that 94\% of the rays could be omitted by integrating foveated rendering with ray tracing. Thus many researchers focused on ray tracing based on foveated rendering with the concept of multi-spatial resolution.


\begin{figure*}[htbp] 
	\centering
	\includegraphics[width=\linewidth]{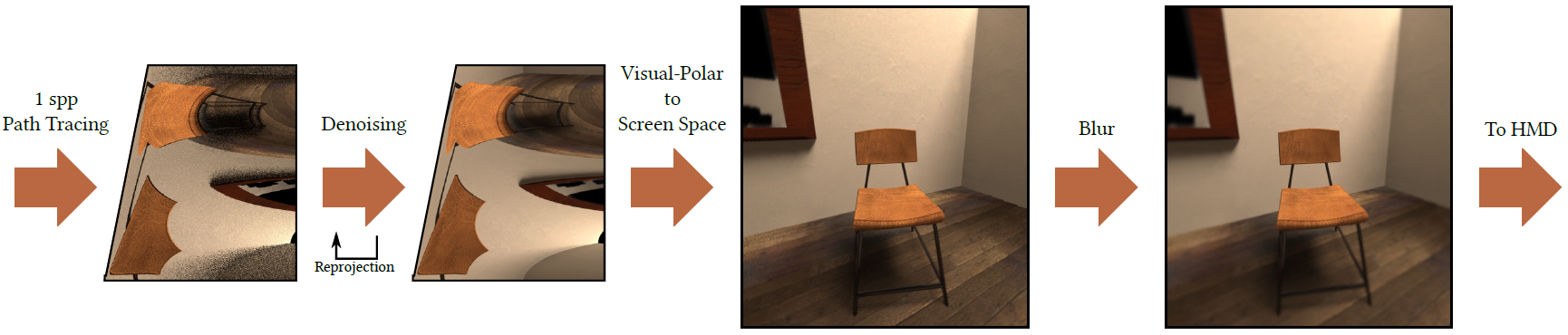}
	\caption{
		{\leftskip=0pt \rightskip=0pt plus 0cm
			Foveated real-time path tracing in visual-polar space proposed by Koskela et al. \cite{koskela2019foveated}.
			Rays were traced and rendering results denoised in a Visual-Polar space, the results were then mapped to the screen space, finally the Guassian blur was performed to generate the final HMD rendering result.
			Ray tracing and denoising in Visual-Polar space  increase both by 2.5$\times$ faster.
			Images courtesy of  Koskela et al. \cite{koskela2019foveated}.}}
	
	\label{fig:rw_2019foveated}
\end{figure*}

Fujita et al. \cite{fujita2014foveated} first implemented the foveated rendering system based on ray tracing. A pre-computed sampling pattern was used with a kNN scheme to reconstruct images from sparse samples. Their system showed artifacts, without considering the eye sensitivity to contrasts and  lacked pertinent input from relevant user studies.
To address these challenges, Weier et al. \cite{weier2016foveated} combined ray tracing based foveated rendering with reprojection rendering, using information from the previous frame to reduce the sampling rays for new frames. Subsequently, the authors applied a temporal caching and resampling scheme to improve reconstruction quality for regions that expose high contrasts and silhouettes. The results of user studies conducted demonstrated that the method achieved a real-time frame rate and compared with the fully rendered image, the visual difference was difficult to detect. 
Blackmon et al. \cite{blackmon2017foveated} combined ray tracing and rasterization in a single pipeline. Ray tracing was used to render the foveal region and rasterization to render the peripheral region.
To speed up previewing the artist’s points of interest, Koskela et al. \cite{koskela2017foveated, koskela2018instantaneous} applied foveated rendering to progressive Monte Carlo rendering, which omits more than 90\% of rays that must be traced in real time. Their user study demonstrated that the perceived convergence of the proposed method was 10$\times$ faster than that of a conventional preview, and participants rated the method to have only marginally more artifacts in areas where it had to start rendering from scratch.
Molenaar et al.  \cite{molenaar2018towards} traced rays based on the visual acuity fall-off model, and reconstructed images based on a spatial CSF. Experimental results demonstrated that this method provided a basic speed improvement of 4.3$\times$.

Willberger et al. \cite{willberger2019deferred} introduced a hybrid path tracing approach to accelerate the global illumination calculation in foveated rendering. The method uses screen space path tracing to render objects with diffuse, specular and glossy materials, using multi-bounced path tracing to render objects with the transparent material.
To render direct lighting from millions of dynamic light sources interactively with ray tracing,
Bitterli et al. \cite{bitterli2020spatiotemporal} introduced the spatiotemporal reservoir resampling method to resample a set of candidate light samples based on the spatio-temporal feature, and subsequently traced rays from sampled lights to illuminate the scene.
Kim et al. \cite{youngwook2021selective} proposed a perceptually efficient pixel sampling method suitable for HMD ray tracing, which combined the  Jin et al. \cite{jin2009selective} selective oversampling technique with the foveated rendering scheme.

\begin{figure*}[htbp]  
	\centering
	\includegraphics[width=\linewidth]{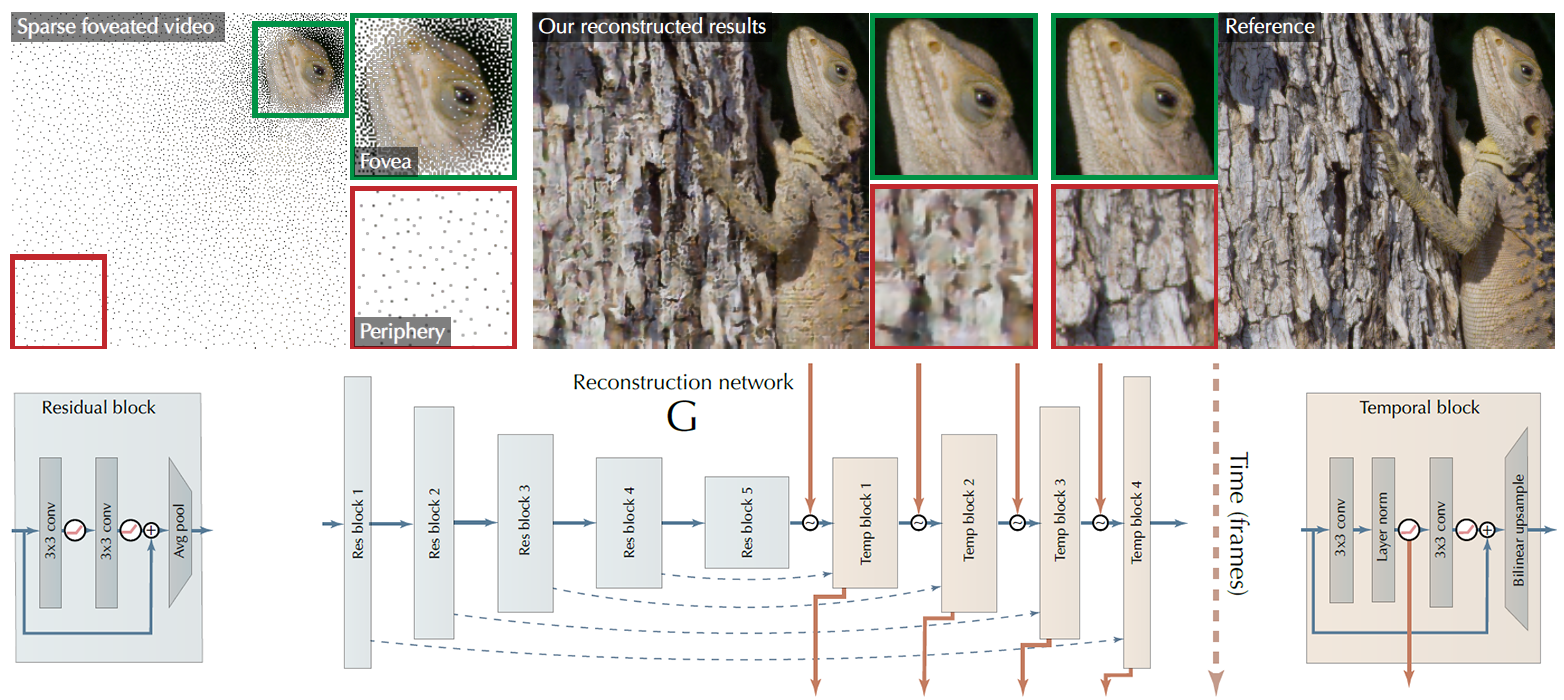}
	\caption{
		{\leftskip=0pt \rightskip=0pt plus 0cm
			Neural reconstruction for foveated rendering and video compression proposed by Kaplanyan et al. \cite{kaplanyan2019deepfovea}. 
			the authors reconstructed the foveated video through a generative adversarial neural network from the sparse foveated video frames with 10\% of pixels (top left). 
			This method reconstructed the video compressed by more than 14$\times$ of the original video, and the reconstructed result (top middle) had no significant reduction in perceptual quality compared with the reference (top right).
			The recurrent video encoder-decoder network architecture is visualized in the bottom. Images courtesy of Kaplanyan et al. \cite{kaplanyan2019deepfovea}.
	}}
	\label{fig:rw_2019deep}
\end{figure*}

As linear falloff still requires many rays in the periphery \cite{fujita2014foveated, weier2016foveated, molenaar2018towards}, Koskela et al. \cite{koskela2019foveated} traced rays and denoised in Visual-Polar space, and subsequently mapped the results to the screen space. In this method, when perceived quality is similar, rendering and denoising speed will increase by 2.5$\times$, and ray traversal speed will increase by 1.3-1.5$\times$. This is because primary rays maintain high coherence, and GPU resource utilization is improved.
The pipeline of this method is shown in Figure \ref{fig:rw_2019foveated}.
Koskela et al. \cite{koskela2020foveated} proposed a working prototype of a foveated ray tracing system that combined the novel Visual-Polar coordinate space proposed in  Koskela et al.  \cite{koskela2019foveated} and the regression-based reconstruction filter proposed in Koskela et al.  \cite{koskela2019Blockwise} for ray tracing that runs in real time.

Most previous methods model the sensitivity as a function of eccentricity and control the number of rays emitted according to these functions, without considering that displayed content also strongly influenced sensitivity. 
Tursun et al. \cite{tursun2019luminance} proposed a new luminance-contrast-aware foveated ray tracing technique. This technique showed that if the spatio-luminance CSF is considered in foveated rendering, the number of tracing rays can be significantly reduced. The disadvantage is that a low-quality image must be generated for each frame, indicating areas with different luminances.

For applying DoF effects in foveated ray tracing, Weier et al. \cite{weier2018foveated} proposed a foveated rendering system that integrates DoF filters to hide potential visual artifacts. Results of the perceptual study showed that tracing rays reduced by more than 69\% while rendering quality of this system was rated almost on par with full rendering.
Liu et al. \cite{jingyu2021perception} developed a mathematical model to simulate the DoF effects of human eyes in VR and subsequently performed DoF-based stochastic sampling to simulate retinal blur according to this mathematical model.

%
%
%
%

\subsection{Muti-spatial Resolution for Image/Video}
\label{sec_recent_MSR_img}


Muti-spatial resolution for image/video research can be divided into three categories: 1) conducting perceptual research on foveated images or videos; 2) neural rendering  on foveated images or videos; 3) accelerating the encoding and transmission of 360$^{\circ}$video streaming.

In the first category, some researchers used high-quality images/videos taken by cameras or rendered with 3D models to generate foveated images/videos by filtering or down-sampling high-quality images/videos in the peripheral region and designed user studies to evaluate foveated rendering performance and quality parameters.
Albert et al. \cite{albert2017latency} explored the effect of foveated rendering latency in VR applications. The results showed that larger foveal regions allow for more aggressive foveation, which is further pronounced for temporally stable foveation techniques. The results also demonstrated that increasing eye-tracking latency by 80–150 ms causes a significant reduction in the acceptable amount of foveation, however, a similar decrease in acceptable foveation was not identified for shorter eye-tracking latencies of 20–40 ms, suggesting that a total system latency of 50–70 ms could be tolerated.
Hsu et al. \cite{hsu2017foveated} proposed a regression model to demonstrate the relationship between human perceived quality and foveated rendering parameters, such as the number of layers, the eccentricity degrees, and resolution of the peripheral region. The results demonstrated that 1) no absolute superior subjective assessment method exists, 2) subjects must complete further observations to confirm that foveated rendering is more imperceptible than perceptible, 3) When the eccentric angle is 7.5$^\circ$ +, and the peripheral region resolution is 540$p$+, subjects barely notice foveated rendering, and 4) the quality of experiments level is highly dependent on the individuals and scenes.

To further improve foveated rendering speed, a small fraction of pixels are provided in the peripheral region for each frame, hence, the image quality of the peripheral region is unacceptable. A neural rendering model was introduced to solve this problem. Kaplanyan et al. \cite{kaplanyan2019deepfovea} proposed a generative adversarial neural network to improve the quality of images/videos in the peripheral region. The method can achieve real-time frame rates with gaze-contingent head-mounted displays on modern hardware. Figure \ref{fig:rw_2019deep} compared the results among the compressed video, reconstructed video, and reference video frames.

\begin{figure*}[htbp]  
	\centering
	\includegraphics[width=\linewidth]{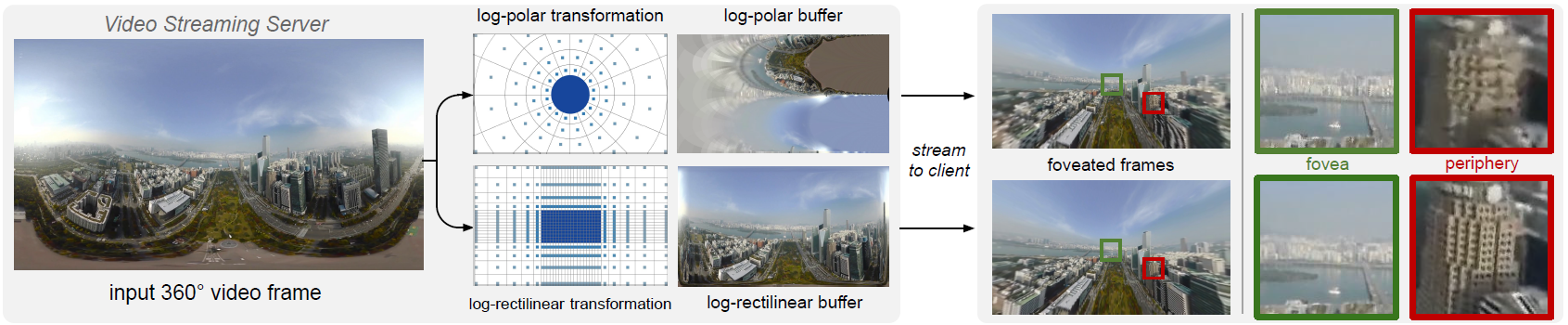}
	\caption{
		{\leftskip=0pt \rightskip=0pt plus 0cm
			Log-rectilinear transformation for foveated 360$^\circ$ video streaming proposed by Li et al. \cite{li2021log}.
			The upper and lower rows present the workflows with prior log-polar transformation and the proposed log-rectilinear transformation respectively. Both foveated methods convert the equirectangular video frames into down-sampled buffers, and subsequently encode and stream buffers to the client.On the client side, buffers are decoded to the screen space to generate the final results. The log-rectilinear transformation reduces flickering and aliasing artifacts in both the foveal and peripheral regions more significantly than that of the prior log-polar transformation.
			Images courtesy of Li et al. \cite{li2021log}.}}
	
	\label{fig:rw_2021log}
\end{figure*}

Some research focuses on improving the accuracy of deep learning models for specific computer vision tasks based on foveated images or videos.
Deza et al. \cite {deza2021emergent} explored the visual representation of the human foveated perceptual system, encoded the feature, and trained a convolutional neural network named Foveation-Nets to perform scene categorization. The results demonstrated that the visual representation of Fovation-Nets learning was different from the network without foveated input, and Fovation-Nets had an impact on generalization, robustness, and perceptual sensitivity. This provided computational support for the idea that the HVS foveated nature may confer a functional advantage for scene representation.
Surace et al. \cite{surace2021learning} proposed a procedure to train a generative network for foveated image reconstruction. This procedure penalized perceptually significant deviations in the output to maintain perceived rather than natural image statistics.

The immersive experience offered in VR via 360$^{\circ}$ video is becoming increasingly popular. However, current bandwidth can barely accommodate the 360$^{\circ}$ video streaming solution that delivers the entire HD 360$^{\circ}$ video frame in real time. As most of the pixels in 360$^{\circ}$ video are invisible or located in peripheral regions, streaming 360$^{\circ}$ video based on the fovea is a more efficient solution. Therefore, encoding and transmission of 360$^{\circ}$ video based on the fovea constitutes important foveated rendering research.
Li et al. \cite{li2021log} proposed a log-linear transformation method to encode original HD 360$^{\circ}$ video frames based on the fovea and to transmit them to HMDs, which maintain full-resolution fidelity in the fovea and have improved perceptual blurring effects in the periphery.
Figure \ref{fig:rw_2021log} compares the final rendering results to the client, encoded by the traditional log-polar transformation and the log-rectilinear transformation in the server, respectively.
To increase the transmission speed of the 360$^{\circ}$ video stream from the server to head-mounted displays, Lungaro et al. \cite{lungaro2018gaze} proposed a gaze-aware transmission approach for 360$^{\circ}$ video streaming services, which delivered high visual quality images around the users’ gaze points in real time while lowering quality elsewhere. The results of user studies demonstrated that compared with traditional solutions, the bandwidth required to provide users with a high quality of experience level, was reduced by up to 83\%.

%
%
%
%

\subsection{LoD}
\label{sec_recent_lod}

In recent years, some research focused on the foveated rendering method based on the LoD technique. 
Different from previous years, researchers focused on designing user studies to optimize or select various parameters involved in the previous method or refine previous methods instead of proposing new LoD methods.

Swafford et al.  \cite{swafford2016user} designed a user study that compares a foveated rendered image with an eccentric angle of 9$^\circ$ and a reference image at full resolution in random order. Three LoDs are generated on the scene geometry: high, medium, and low. A lower level means that there is a less tessellated grid for each tile. The results demonstrated that users had a similar visual experience to the foveated LoD rendered image with the medium level in the peripheral region and the full-resolution reference image. However, time performance could be improved by 3$\times$.
As Swafford et al.  \cite{swafford2016user} only applied the tessellation method to fixed-size triangles, the results of tessellation of much larger or smaller triangles do not match the visual perceptual size.
Zheng et al. \cite{zheng2018perceptual} adaptively adjusted the tessellation levels and culling region based on visual sensitivity.
Young et al. \cite{young2019real} adjusted the foveal region size and shape to correct the gaze tracing error or state parameters and combined this technique with LoD to render foveated images.
Stafford et al. \cite{stafford2019selective} selectively filtered the images in the peripheral region to reduce visual artifacts owing to contrast resulting from the lower LoD before compositing foveated images for presentation.
Lindeberg et al.  \cite{lindeberg2016concealing} proposed a gaze-contingent depth of field tessellation that applies tessellation to all objects within the focal plane, gradually decreasing tessellation levels as applied blur increases. 
User studies demonstrated that this technique helps reduce the number of primitives rendered by approximately 70\% and frame times by approximately 9\% compared with using fully adaptive tessellation.

Researchers not only applied LoD-based foveated rendering to scenes with geometric meshes, but also to point clouds to improve time performance. Schutz et al. \cite{schutz2019real} proposed a continuous LoD method for rendering large point clouds in real time. This method continuously recreated a down-sampled vertex buffer from the full point cloud, based on camera orientation, position, and distance to the camera, in a point-wise fashion and at a speed of 17 million points per millisecond.


\subsection{Multi-spatial Resolution for Volume Data} \label{sec:vol}

In recent years, with the increase in GPU computing power, researchers have further proposed more complex techniques to improve the efficiency of volume data foveated rendering.

Gallo et al. \cite{gallo2013high} introduced a hybrid CPU-GPU volume ray-casting system for interactive, medical-quality visualization using an ordinary desktop PC. The system combined three parts: a gaze-directed volume rendering tool that renders the foveal region in maximum resolution, an inner structure tool that enables interactive inspection of data using two different transfer functions simultaneously, and a localized oversampling tool that allows users to interactively execute oversampling and antialiasing techniques in the foveal region.
Bruder et al. \cite{bruder2019voronoi} accelerated volume rendering through the Linde-Buzo-Gray sampling method based on the visual acuity fall-off model and natural neighbor interpolation.
Ananpiriyakul et al. \cite{ananpiriyakul2020gaze} smoothly transited the resolution from the foveal to the peripheral region with the use of face-tracking to drive adaptive-resolution volume data visualization. The results demonstrated a 2-2.5$\times$ frame rate improvement on interactive explorations.
Kang et al. \cite{kang2020depth} proposed a thin lens camera model to simulate rays passing through different parts of the lens for volume data visualizations. The model is implemented in the GPU pipeline with no pre-processing. The results demonstrated that the method could generate volume data visualizations with better depth perception than existing DoF methods, and the speed was 9$\times$ faster.

\subsection{Multi-luminance Resolution} \label{sec: mol}
The concept of multi-luminance resolution has only been used in foveated rendering in the past 5 years.

\begin{figure}[htbp] 
	\centering
	\includegraphics[width=\linewidth]{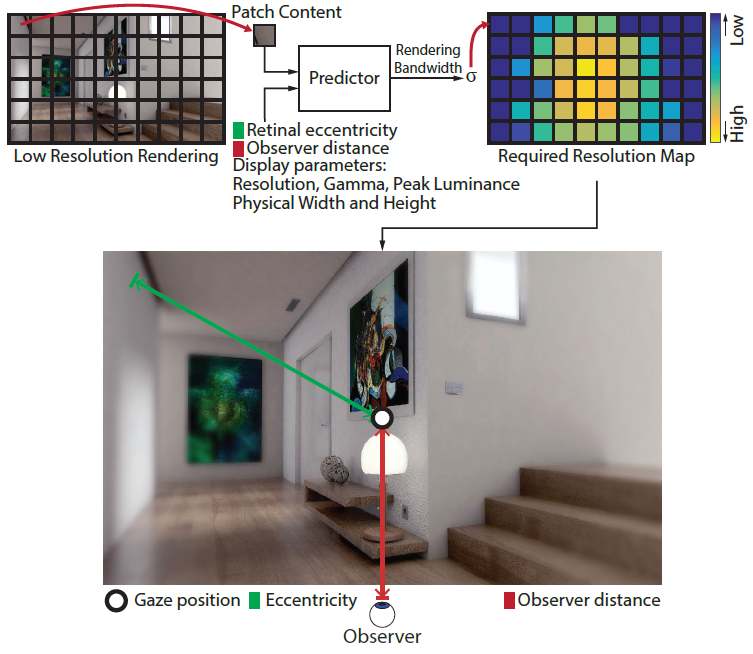}
	\caption{
		{\leftskip=0pt \rightskip=0pt plus 0cm
			Luminance-contrast-aware foveated rendering proposed by Tursun et al. \cite{tursun2019luminance}. 
			A low-resolution image was first rendered, and then divided into multiple small patches, subsequently, the standard deviation $\rho$ was calculated to obtain the maximum acceptable resolution reduction for each patch. Finally, the luminance-contrast-aware adaptive resolution rendering was performed through real-time ray tracing. Compared with standard foveated rendering, this method achieved a 0.8-2.6$\times$ acceleration and improved perceptual quality.
			Images courtesy of Tursun et al. \cite{tursun2019luminance}.
		}
	}
	\label{fig:rw_2019luminance}
\end{figure}

Stengel et al. \cite{stengel2016adaptive} presented a luminance map to adjust the sampling probability of the periphery to obtain shading samples that can effectively shade important features of the image.
Tursun et al. \cite{tursun2019luminance} proposed a novel luminance-contrast-aware foveated rendering technique that improves computational savings by analyzing the local luminance contrast of the image, this method pipeline is demonstrated in Figure \ref{fig:rw_2019luminance}.
Wang et al. \cite{wang2020foveated} proposed the foveated instant radiosity method that casts more VPLs to illuminate the foveal region such that more accurate global illumination effects in the foveal region and less accurate global illumination in the peripheral region can be rendered.
Yang et al. \cite{yang2021foveated} improved the method proposed by Wang et al. \cite{wang2020foveated} and created a CMF-based perceptual probability map to manage virtual point lights more accurately to further improve rendering quality in the fovea.
Because the method of Wang et al. \cite{wang2020foveated} and Yang et al. \cite{yang2021foveated} only supports diffuse scenes, Shi et al. \cite{shi2021foveated} adopted the photon mapping method to foveated rendering, which renders high-quality global illumination effects in the foveal region at interactive frame rates for the scenes that include diffuse, specular, glossy and transparent materials.


\subsection{Foveated Rendering for Nascent Data Types}
\label{sec_recent_MSR_other}

With the rise of 3D display technologies, new data types appear, such as hologram data and light fields. However, current hardware and graphic algorithms cannot enable high quality and low latency for 3D displays. Researchers extend the foveated rendering algorithms to support these nascent data types.

Researchers extended foveated rendering methods from 3D geometry scenes to 4D light fields based on the concept of multi-spatial resolution.
Sun et al. \cite{sun2017perceptually} proposed a 4D light field foveated rendering method with importance sampling and a sparse reconstruction scheme based on the spectral bounds and depth perception measurements. The results demonstrated that the technique traced only 16-30\% rays without compromising perceptual quality.
Meng et al. \cite{meng20203d} introduced a 3D-kernel foveated rendering method to observe light fields, which provided similar visual results as the original light fields. However, this achieves a speed improvement of up to 7.28$\times$ for the light fields with a resolution of 25$\times$25$\times$1024$\times$1024$p$ with minimal perceptual loss of detail.

Foveated rendering research has also been published based on the concept of multi-spatial resolution to improve the rendering of holograms.
Wei et al. \cite{wei2019fast} proposed an angle-changeable foveated ray tracing method for rendering the computer-generated hologram (CGH) with better performance and almost no observable artifacts for the user. 
Chakravarthula et al. \cite{chakravarthula2021gaze}
reduced the perceived speckle noise by integrating two factors into the phase hologram computation: 1) foveal and peripheral vision HVS characteristics; 2) the retinal point spread function. With this new method, the perceived speckle noise can be pushed from the fovea to the periphery.


\section{Discussion} \label{sec_OOQ}
Although foveated rendering has been a focus area in research and industry for more than two decades, there are still many opportunities and open questions to be solved.


One potential opportunity is to take full advantage of the human visual features for foveated rendering. The current foveated rendering method only uses parts of the HVS features, including visual acuity and contrast sensitivity, and other features that may be beneficial in this context are not reflected in existing research, therefore, further research is required to investigate this. For example, visual masking may be utilized for accelerating foveated rendering. This explains that the visibility of one image, called a target, can be reduced by the presence of another image, called a mask \cite{sherrington1897onReciprocal}. For example, as the luminance or scene changes sharply, the HVS sensitivity will decrease when a new scene suddenly appears. Therefore, decreasing rendering quality of the foveal image in the subsequent frames will not cause the user to notice the difference. We believe that the next important step towards foveated rendering is effectively capitalizing of human visual features to achieve more aggressive foveated rendering without compromising perceptual awareness.



Another potential opportunity is to apply computer vision and artificial intelligence technologies to address some issues for current foveated rendering methods. 
Some explorations on this aspect have been completed. 
To further improve user gaze tracking accuracy, Arabadzhiyska et al. \cite{arabadzhiyska2017saccade} proposed a method to predict the landing position of the gaze position during saccades in foveated rendering preprocessing. Kaplanyan et al. \cite{kaplanyan2019deepfovea} employed a generated adversarial neural network in the foveated rendering post-processing stage, which reconstructed details in the fovea and generated temporally stable peripheral content. Other technologies, for example, the attention model, could also be considered for integration into the foveated rendering paradigm to improve quality and performance.

The development of cutting-edge foveated displays is another potential avenue for foveated rendering. In recent years, Tan et al. \cite{tan2018foveated} used beam splitters with different magnifications to combine two identical displays to demonstrate a dynamic foveal VR display. Lee et al. \cite{lee2019enhanced} introduced a time-multiplexed see-through fixed foveated holographic display using a beam splitter and tunable lens, with a foveal field of view of 1.04$^\circ$ and a peripheral field of view of 22.6$^\circ$. Kim et al. \cite{kim2019foveated} presented a foveated display with resolution and focal depth dynamically driven by gaze tracking for AR. The display combines a traveling micro-display for the high-resolution foveal region with a wide field-of-view peripheral display that follows the viewer's pupil during eye movement. However, current foveated displays for VR and AR have high mechanical complexities and drawbacks for responsiveness and power draw. Focus depth estimation of current displays is not robust, although previous research supports the feasibility of estimating focal depth based on binocular astigmatism alone, it has also been reported that half diopters or more are inaccurate \cite{mlot20163d}. The combination of foveated displays and prescription corrective optics also presents a challenge.

Based on the analysis and summary of existing foveated rendering methods, some open questions require urgent solutions.

Currently, many studies have been published on foveated rendering methods for volume data and geometric meshes, and concepts are relatively mature. Only in recent years foveated rendering research of hologram data and the light fields is nascent. Generally, foveated rendering methods involving volume data and geometric meshes are used for reference, such as the ray tracing method. Thus, further research is required to identify a more suitable foveated rendering method for these new data types.

Although the ray tracing framework can be adopted into foveated rendering in a straightforward manner, this is inefficient for some special effects in 3D rendering, such as global illumination for the scene containing point light sources, and high-detailed caustics. Some rendering paradigms render these special effects more efficiently, however, they cannot be directly integrated into foveated rendering. Adopting these rendering paradigms to support foveated rendering is therefore a challenge. Methods proposed in \cite{wang2020foveated, yang2021foveated, shi2021foveated} are interesting attempts. Based on the concept of multi-luminance resolution, they adopt instant radiosity and photon mapping to foveated rendering. Based on different foveation principles, many other efficient real-time rendering paradigms, such as bidirectional path tracing \cite{veach1997robust} and vertex connection and merging \cite{georgiev2012light}, etc., can be applied to foveated rendering for improved performance.

To evaluate foveated images/video quality, the straightforward method is to design perceptual experiments to collect user's perception information. As perceptual experiments are typically time-consuming and costly, they should be performed for methods with a greater chance of success. Therefore, some objective metrics based on the biological and physical theories involving foveated rendering must be proposed to quickly evaluate the feasibility of the tested foveated rendering methods.
Currently, some metrics exist to evaluate foveated image quality, for example:
(1) The foveal signal-to-noise ratio (FSNR) \cite{lee2002foveated} valued the distortion between foveated images and reference images with a weighted signal-to-noise ratio. FSNR failed to consider user perception of foveated images quality, which may cause perceptual deviations in evaluating foveated images.
(2) The foveated wavelet image quality index (FWQI) \cite{wang2003foveated} calculated the wavelet coefficient difference between foveated and reference images with the integration of spatial CSF. FWQI did not consider spatio-temporal CSF while it was reported that the contrast sensitivity of the HVS can be significantly influenced by the retinal velocity \cite{you2013attention}.
(3) The foveated mean squared error (FMSE) \cite{rimac2010foveated} evaluated foveated video quality with the consideration of both spatial and spatio-temporal CSF. FMSE assumed that eye fixation points are always located at the center of images. This assumption potentially introduces biases in evaluating visual quality.
(4) The window-based structural similarity index (WSSIM) \cite{tsai2014foveation} used different rules to evaluate foveated image quality for different windows on the foveated images, the scoring rules for the window closer to the fovea will be more stringent. WSSIM relies on selecting an appropriate saliency model. However, this may bias foveated image evaluation results.
Thus far, the lack of a more general, comprehensive, and widely accepted metric has significantly complicated the evaluation of foveated images/video quality. In addition, constructing datasets to evaluate different foveated images/video aspects could ensure improved comparability of evaluation results.

In recent years, most foveated rendering methods designed are for VR applications, and few methods aim toward AR applications.
Kim et al. \cite{kim2019foveated} investigated foveated rendering under AR. The focus was predominantly on the design of a dynamically-foveated augmented reality display. For AR applications that require virtual and real fusion, the degree of fusion will directly affect the quality of rendering results, therefore the question of how to control the degree of fusion to generate images of different qualities in different regions remains an open challenge. For information-enhanced AR applications, it is also worth exploring whether relevant content such as scene semantic and task target information can be added to foveated rendering.

In addition to improving rendering speed, foveated rendering can also be used to complete specific tasks. For example, Joshi et al. \cite{joshi2020inattentional} presented foveated rendering-based redirected walking in VR, which rendered a high-quality region to guide the spatially-varying rotation and updated peripheral framebuffer during inattentional blindness. Whether foveated rendering can assist or improve other VR and AR tasks is yet to be explored.

\section{Conclusion}
This paper surveys research and development involving foveated rendering over the past 31 years. Visual perception theories and taxonomies regarding foveated rendering are discussed in-depth. We respectively review early foveated rendering technologies (from 1990 to 2011) and those that have more recently emerged over the past decade (from 2012 to 2021) Finally, we discuss potential opportunities and open questions for future research in this field.

\appendix

\subsection*{Acknowledgements}
This work was supported in part by the National Key R\&D Plan 2019YFC1521102, by the National Natural Science Foundation of China through Projects 61932003 and 61772051, by the Beijing Natural Science Foundation L182016, by the Beijing Program for International S\&T Cooperation Project Z191100001619003, by the funding of Shenzhen Research Institute of Big Data (Shenzhen 518000).

\subsection*{Declaration of competing interest}
The authors have no competing interests to declare that are relevant to the
content of this article.
 \bibliographystyle{CVMbib}
\bibliography{fr_ref}

\begin{thebibliography}{100}
\expandafter\ifx\csname urlstyle\endcsname\relax
  \providecommand{\doi}[1]{doi:\discretionary{}{}{}#1}\else
  \providecommand{\doi}{doi:\discretionary{}{}{}\begingroup
  \urlstyle{rm}\Url}\fi

\bibitem{correa2009evaluation}
Corr{\^e}a CG, Nunes FL, Bezerra A, Carvalho~Jr PM. Evaluation of VR medical
  training applications under the focus of professionals of the health area. In
  \emph{Proceedings of the 2009 ACM symposium on Applied Computing}, 2009,
  821--825.

\bibitem{hsieh2017vr}
Hsieh MC, Lin YH. VR and AR applications in medical practice and education.
  \emph{Hu Li Za Zhi}, 2017, 64(6): 12--18.

\bibitem{hsieh2018preliminary}
Hsieh MC, Lee J. Preliminary study of VR and AR applications in medical and
  healthcare education. \emph{J Nurs Health Stud}, 2018, 3(1): 1.

\bibitem{rizzo2005human}
Rizzo A, Morie JF, Williams J, Pair J, Buckwalter JG. Human emotional state and
  its relevance for military VR training. Technical report, University Of
  Southern California Marina Del Rey Ca Inst For Creative~…, 2005.

\bibitem{lele2013virtual}
Lele A. Virtual reality and its military utility. \emph{Journal of Ambient
  Intelligence and Humanized Computing}, 2013, 4(1): 17--26.

\bibitem{ahir2020application}
Ahir K, Govani K, Gajera R, Shah M. Application on virtual reality for enhanced
  education learning, military training and sports. \emph{Augmented Human
  Research}, 2020, 5(1): 1--9.

\bibitem{ong2013virtual}
Ong SK, Nee AYC.  \emph{Virtual and augmented reality applications in
  manufacturing}, Springer Science \& Business Media2013.

\bibitem{choi2015virtual}
Choi S, Jung K, Noh SD. Virtual reality applications in manufacturing
  industries: Past research, present findings, and future directions.
  \emph{Concurrent Engineering}, 2015, 23(1): 40--63.

\bibitem{doolani2020review}
Doolani S, Wessels C, Kanal V, Sevastopoulos C, Jaiswal A, Nambiappan H,
  Makedon F. A review of extended reality (xr) technologies for manufacturing
  training. \emph{Technologies}, 2020, 8(4): 77.

\bibitem{avila2014virtual}
Avila L, Bailey M. Virtual reality for the masses. \emph{IEEE computer graphics
  and applications}, 2014, 34(05): 103--104.

\bibitem{bialkova2017sound}
Bialkova S, Van~Gisbergen MS. When sound modulates vision: VR applications for
  art and entertainment. In \emph{2017 IEEE 3rd Workshop on Everyday Virtual
  Reality (WEVR)}, IEEE2017, 1--6.

\bibitem{saint2021survey}
Saint-Louis C, Hamam A. Survey of Haptic Technology and Entertainment
  Applications. In \emph{SoutheastCon 2021}, IEEE2021, 01--07.

\bibitem{puggioni2020content}
Puggioni MP, Frontoni E, Paolanti M, Pierdicca R, Malinverni ES, Sasso M. A
  content creation tool for AR/VR applications in education: The ScoolAR
  framework. In \emph{International conference on augmented reality, virtual
  reality and computer graphics}, Springer2020, 205--219.

\bibitem{ferdani20203d}
Ferdani D, Fanini B, Piccioli MC, Carboni F, Vigliarolo P. 3D reconstruction
  and validation of historical background for immersive VR applications and
  games: The case study of the Forum of Augustus in Rome. \emph{Journal of
  Cultural Heritage}, 2020, 43: 129--143.

\bibitem{tanenbaum2020make}
Tanenbaum TJ, Hartoonian N, Bryan J. " How do I make this thing smile?" An
  Inventory of Expressive Nonverbal Communication in Commercial Social Virtual
  Reality Platforms. In \emph{Proceedings of the 2020 CHI Conference on Human
  Factors in Computing Systems}, 2020, 1--13.

\bibitem{potter2014detecting}
Potter MC, Wyble B, Hagmann CE, McCourt ES. Detecting meaning in RSVP at 13 ms
  per picture. \emph{Attention, Perception, \& Psychophysics}, 2014, 76(2):
  270--279.

\bibitem{hendrickson1984morphological}
Hendrickson AE, Yuodelis C. The morphological development of the human fovea.
  \emph{Ophthalmology}, 1984, 91(6): 603--612.

\bibitem{loschky2001perceptual}
Loschky LC, McConkie GW, Yang J, Miller ME. Perceptual effects of a
  gaze-contingent multi-resolution display based on a model of visual
  sensitivity. In \emph{the ARL Federated Laboratory 5th Annual Symposium-ADID
  Consortium Proceedings}, 2001, 53--58.

\bibitem{luebke2001perceptually}
Luebke D, Hallen B. Perceptually driven simplification for interactive
  rendering. In \emph{Eurographics Workshop on Rendering Techniques},
  Springer2001, 223--234.

\bibitem{zheng2018perceptual}
Zheng Z, Yang Z, Zhan Y, Li Y, Yu W. Perceptual model optimized efficient
  foveated rendering. In \emph{Proceedings of the 24th ACM Symposium on Virtual
  Reality Software and Technology}, 2018, 1--2.

\bibitem{schutz2019real}
Sch{\"u}tz M, Kr{\"o}sl K, Wimmer M. Real-time continuous level of detail
  rendering of point clouds. In \emph{2019 IEEE Conference on Virtual Reality
  and 3D User Interfaces (VR)}, IEEE2019, 103--110.

\bibitem{loschky2000user}
Loschky LC, McConkie GW. User performance with gaze contingent
  multiresolutional displays. In \emph{Proceedings of the 2000 symposium on Eye
  tracking research \& applications}, 2000, 97--103.

\bibitem{parkhurst2002variable}
Parkhurst DJ, Niebur E. Variable-resolution displays: A theoretical, practical,
  and behavioral evaluation. \emph{Human factors}, 2002, 44(4): 611.

\bibitem{duchowski2014reducing}
Duchowski AT, House DH, Gestring J, Wang RI, Krejtz K, Krejtz I, Mantiuk R,
  Bazyluk B. Reducing visual discomfort of 3D stereoscopic displays with
  gaze-contingent depth-of-field. In \emph{Proceedings of the acm symposium on
  applied perception}, 2014, 39--46.

\bibitem{turner2018phase}
Turner E, Jiang H, Saint-Macary D, Bastani B. Phase-aligned foveated rendering
  for virtual reality headsets. In \emph{2018 IEEE Conference on Virtual
  Reality and 3D User Interfaces (VR)}, IEEE2018, 1--2.

\bibitem{guenter2012foveated}
Guenter B, Finch M, Drucker S, Tan D, Snyder J. Foveated 3D graphics. \emph{ACM
  Transactions on Graphics (TOG)}, 2012, 31(6): 1--10.

\bibitem{bastani2020smoothly}
Bastani B, Funt B, Vignaud S, Jiang H. Smoothly varying foveated rendering,
  2020, uS Patent 10,546,364.

\bibitem{stengel2016adaptive}
Stengel M, Grogorick S, Eisemann M, Magnor M. Adaptive Image-Space Sampling for
  Gaze-Contingent Real-time Rendering. \emph{Computer Graphics Forum}, 2016,
  35: 129--139, \doi{https://doi.org/10.1111/cgf.12956}.

\bibitem{tursun2019luminance}
Tursun OT, Arabadzhiyska-Koleva E, Wernikowski M, Mantiuk R, Seidel HP,
  Myszkowski K, Didyk P. Luminance-contrast-aware foveated rendering. \emph{ACM
  Transactions on Graphics (TOG)}, 2019, 38(4): 1--14.

\bibitem{tavakoli2019scene}
Tavakoli M, Khan M, Renschler M, Mondal M. Scene-based foveated rendering of
  graphics content, 2019, uS Patent 10,482,648.

\bibitem{koskela2016foveated}
Koskela M, Viitanen T, J{\"a}{\"a}skel{\"a}inen P, Takala J. Foveated Path
  Tracing. In G~Bebis, R~Boyle, B~Parvin, D~Koracin, F~Porikli, S~Skaff,
  A~Entezari, J~Min, D~Iwai, A~Sadagic, C~Scheidegger, T~Isenberg, editors,
  \emph{Advances in Visual Computing}, Cham: Springer International
  Publishing2016, 723--732.

\bibitem{molenaar2018towards}
Molenaar EN. Towards real-time ray tracing through foveated rendering. Master's
  thesis, Utrecht University, 2018.

\bibitem{koskela2019foveated}
Koskela M, Lotvonen A, Mäkitalo M, Kivi P, Viitanen T, Jääskeläinen P.
  Foveated Real-Time Path Tracing in Visual-Polar Space. In T~Boubekeur, P~Sen,
  editors, \emph{Eurographics Symposium on Rendering - DL-only and Industry
  Track}, The Eurographics Association2019, \doi{10.2312/sr.20191219}.

\bibitem{koskela2020foveated}
Koskela M. \emph{Foveated Path Tracing with Fast Reconstruction and Efficient
  Sample Distribution}. Tampere University Dissertations - Tampereen yliopiston
  v{\"a}it{\"o}skirjat, Tampere University2020.

\bibitem{levoy1990gaze}
Levoy M, Whitaker R. Gaze-directed volume rendering. In \emph{Proceedings of
  the 1990 symposium on Interactive 3D graphics}, 1990, 217--223.

\bibitem{wang2020foveated}
Wang L, Li R, Shi X, Yan LQ, Li Z. Foveated Instant Radiosity. In \emph{2020
  IEEE International Symposium on Mixed and Augmented Reality (ISMAR)},
  IEEE2020, 1--11.

\bibitem{bruder2019voronoi}
Bruder V, Schulz C, Bauer R, Frey S, Weiskopf D, Ertl T. Voronoi-Based Foveated
  Volume Rendering. In \emph{EuroVis (Short Papers)}, 2019, 67--71.

\bibitem{kaplanyan2019deepfovea}
Kaplanyan AS, Sochenov A, Leimk{\"u}hler T, Okunev M, Goodall T, Rufo G.
  DeepFovea: Neural reconstruction for foveated rendering and video compression
  using learned statistics of natural videos. \emph{ACM Transactions on
  Graphics (TOG)}, 2019, 38(6): 1--13.

\bibitem{weier2017perception}
Weier M, Stengel M, Roth T, Didyk P, Eisemann E, Eisemann M, Grogorick S,
  Hinkenjann A, Kruijff E, Magnor M, et~al.. Perception-driven accelerated
  rendering. In \emph{Computer Graphics Forum}, volume~36, Wiley Online
  Library2017, 611--643.

\bibitem{cline1980dictionary}
Cline D.  \emph{Dictionary of visual science}, Chilton Book Company1980.

\bibitem{ivanvcic2019impact}
Ivan{\v{c}}i{\'c}~Valenko S, Cvilju{\v{s}}ac V, Zlati{\'c} S, Modri{\'c} D. The
  Impact of Physical Parameters on the Perception of the Moving Elements in
  Peripheral Part of the Screen. \emph{Tehni{\v{c}}ki vjesnik}, 2019, 26(5):
  1444--1450.

\bibitem{schaadt2015disorders}
Schaadt AK. Disorders of binocular convergent fusion and stereoscopic space
  perception following acquired brain damage: treatment and neuroanatomical
  implications, 2015.

\bibitem{strasburger2011peripheral}
Strasburger H, Rentschler I, J{\"u}ttner M. Peripheral vision and pattern
  recognition: A review. \emph{Journal of vision}, 2011, 11(5): 13--13.

\bibitem{fender1967extension}
Fender D, Julesz B. Extension of Panum's fusional area in binocularly
  stabilized vision. \emph{J.opt.soc.am}, 1967, 57(6): 819.

\bibitem{georgeson2014binocular}
Georgeson MA, Wallis SA. Binocular fusion, suppression and diplopia for blurred
  edges. \emph{Ophthalmic and Physiological Optics}, 2014, 34(2): 163--185.

\bibitem{porac1976dominant}
Porac C, Coren S. The dominant eye. \emph{Psychological bulletin}, 1976, 83(5):
  880.

\bibitem{robson1966spatial}
Robson JG. Spatial and temporal contrast-sensitivity functions of the visual
  system. \emph{Josa}, 1966, 56(8): 1141--1142.

\bibitem{campbell1968application}
Campbell FW, Robson JG. Application of Fourier analysis to the visibility of
  gratings. \emph{The Journal of physiology}, 1968, 197(3): 551.

\bibitem{kelly1979motion}
Kelly D. Motion and vision. II. Stabilized spatio-temporal threshold surface.
  \emph{Josa}, 1979, 69(10): 1340--1349.

\bibitem{mullen1985contrast}
Mullen KT. The contrast sensitivity of human colour vision to red-green and
  blue-yellow chromatic gratings. \emph{The Journal of physiology}, 1985,
  359(1): 381--400.

\bibitem{geisler1998real}
Geisler WS, Perry JS. Real-time foveated multiresolution system for
  low-bandwidth video communication. In \emph{Human vision and electronic
  imaging III}, volume 3299, International Society for Optics and
  Photonics1998, 294--305.

\bibitem{krajancich2021perceptual}
Krajancich B, Kellnhofer P, Wetzstein G. A Perceptual Model for
  Eccentricity-dependent Spatio-temporal Flicker Fusion and its Applications to
  Foveated Graphics. \emph{arXiv preprint arXiv:2104.13514}, 2021.

\bibitem{weymouth1958visual}
Weymouth FW. Visual sensory units and the minimal angle of resolution.
  \emph{American journal of ophthalmology}, 1958, 46(1): 102--113.

\bibitem{weymouth1963visual}
Weymouth FW. Visual sensory units and the minimum angle of resolution.
  \emph{Optometry and Vision Science}, 1963, 40(9): 550--568.

\bibitem{daniel1961representation}
Daniel P, Whitteridge D. The representation of the visual field on the cerebral
  cortex in monkeys. \emph{The Journal of physiology}, 1961, 159(2): 203--221.

\bibitem{levi1985vernier}
Levi DM, Klein SA, Aitsebaomo A. Vernier acuity, crowding and cortical
  magnification. \emph{Vision research}, 1985, 25(7): 963--977.

\bibitem{nakayama1990properties}
Nakayama K. Properties of early motion processing: Implications for the sensing
  of egomotion. \emph{The Perception and Control of Self Motion}, 1990: 69--80.

\bibitem{ohshima1996gaze}
Ohshima T, Yamamoto H, Tamura H. Gaze-directed adaptive rendering for
  interacting with virtual space. In \emph{Proceedings of the IEEE 1996 Virtual
  Reality Annual International Symposium}, IEEE1996, 103--110.

\bibitem{luebke2000perceptually}
Luebke D, Hallen B, Newfield D, Watson B. Perceptually driven simplification
  using gaze-directed rendering. Technical report, Tech. Rep. CS-2000-04,
  Department of Computer Science, University of~…, 2000.

\bibitem{parkhurst2001evaluating}
Parkhurst D, Law I, Niebur E. Evaluating gaze-contingent level of detail
  rendering of virtual environments using visual search, 2001.

\bibitem{vaidyanathan2014coarse}
Vaidyanathan K, Salvi M, Toth R, Foley T, Akenine-M{\"o}ller T, Nilsson J,
  Munkberg J, Hasselgren J, Sugihara M, Clarberg P, et~al.. Coarse pixel
  shading. In \emph{Proceedings of High Performance Graphics}, 2014, 9--18.

\bibitem{weier2016foveated}
Weier M, Roth T, Kruijff E, Hinkenjann A, Pérard-Gayot A, Slusallek P, Li Y.
  {Foveated Real-Time Ray Tracing for Head-Mounted Displays}. \emph{Computer
  Graphics Forum}, 2016, \doi{10.1111/cgf.13026}.

\bibitem{mikkola2010relative}
Mikkola M, Boev A, Gotchev A. Relative Importance of Depth Cues on Portable
  Autostereoscopic Display. In \emph{Proceedings of the 3rd Workshop on Mobile
  Video Delivery}, MoViD '10, New York, NY, USA: Association for Computing
  Machinery2010, 63–68, \doi{10.1145/1878022.1878038}.

\bibitem{panum1858physiologische}
Panum PL.  \emph{Physiologische Untersuchungen {\"u}ber das Sehen mit zwei
  Augen}, Schwer1858.

\bibitem{mitchell1966review}
Mitchell D. A review of the concept of “Panum's fusional areas”.
  \emph{Optometry and Vision Science}, 1966, 43(6): 387--401.

\bibitem{hillaire2008using}
Hillaire S, L{\'e}cuyer A, Cozot R, Casiez G. Using an eye-tracking system to
  improve camera motions and depth-of-field blur effects in virtual
  environments. In \emph{2008 IEEE virtual reality conference}, IEEE2008,
  47--50.

\bibitem{mantiuk2011gaze}
Mantiuk R, Bazyluk B, Tomaszewska A. Gaze-dependent depth-of-field effect
  rendering in virtual environments. In \emph{International Conference on
  Serious Games Development and Applications}, Springer2011, 1--12.

\bibitem{2014Gaze}
Vinnikov M, Allison RS. Gaze-contingent depth of field in realistic scenes: the
  user experience. In \emph{ACM}, 2014, 119--126.

\bibitem{mauderer2014depth}
Mauderer M, Conte S, Nacenta MA, Vishwanath D. Depth perception with
  gaze-contingent depth of field. In \emph{Proceedings of the SIGCHI Conference
  on Human Factors in Computing Systems}, 2014, 217--226.

\bibitem{kusha2016Gaze}
Kushagr G, Suleman K. Gaze Contingent Depth Of Field Display, 2016, 35(6): 179.

\bibitem{weier2018foveated}
Weier M, Roth T, Hinkenjann A, Slusallek P. Foveated depth-of-field filtering
  in head-mounted displays. \emph{ACM Transactions on Applied Perception
  (TAP)}, 2018, 15(4): 1--14.

\bibitem{kang2020depth}
Kang J, Lee J, Shin YG, Kim B. Depth-of-Field Rendering Using Progressive Lens
  Sampling in Direct Volume Rendering. \emph{IEEE Access}, 2020, 8:
  93335--93345.

\bibitem{shneor2006eye}
Shneor E, Hochstein S. Eye dominance effects in feature search. \emph{Vision
  Research}, 2006, 46(25): 4258--4269.

\bibitem{kocctekin2013relation}
Ko{\c{c}}tekin B, G{\"u}ndo{\u{g}}an N{\"U}, Alt{\i}nta{\c{s}} AGK,
  Yaz{\i}c{\i} AC. Relation of eye dominancy with color vision discrimination
  performance ability in normal subjects. \emph{International journal of
  ophthalmology}, 2013, 6(5): 733.

\bibitem{meng2020eye}
Meng X, Du R, Varshney A. Eye-dominance-guided Foveated Rendering. \emph{IEEE
  Transactions on Visualization and Computer Graphics}, 2020, 26(5):
  1972--1980.

\bibitem{owsley2003contrast}
Owsley C. Contrast sensitivity. \emph{Ophthalmology Clinics of North America},
  2003, 16(2): 171--177.

\bibitem{2013Measurements}
Kim KJ, Mantiuk R, Lee KH. Measurements of achromatic and chromatic contrast
  sensitivity functions for an extended range of adaptation luminance.
  \emph{Proceedings of SPIE - The International Society for Optical
  Engineering}, 2013, 8651: 86511A--86511A--14.

\bibitem{chwesiuk2019measurements}
Chwesiuk M, Mantiuk R. Measurements of contrast sensitivity for peripheral
  vision. In \emph{ACM Symposium on Applied Perception 2019}, 2019, 1--9.

\bibitem{tyler1990analysis}
Tyler CW, Hamer RD. Analysis of visual modulation sensitivity. IV. Validity of
  the Ferry--Porter law. \emph{JOSA A}, 1990, 7(4): 743--758.

\bibitem{watson2000vis}
Watson AB. Visual detection of spatial contrast patterns: Evaluation of five
  simple models. \emph{Optics Express}, 2000, 6(1): 12--33.

\bibitem{yee2001spatiotemporal}
Yee H, Pattanaik S, Greenberg DP. Spatiotemporal sensitivity and visual
  attention for efficient rendering of dynamic environments. \emph{ACM
  Transactions on Graphics (TOG)}, 2001, 20(1): 39--65.

\bibitem{westland2006model}
Westland S, Owens H, Cheung V, Paterson-Stephens I. Model of luminance
  contrast-sensitivity function for application to image assessment.
  \emph{Color Research \& Application: Endorsed by Inter-Society Color Council,
  The Colour Group (Great Britain), Canadian Society for Color, Color Science
  Association of Japan, Dutch Society for the Study of Color, The Swedish
  Colour Centre Foundation, Colour Society of Australia, Centre Fran{\c{c}}ais
  de la Couleur}, 2006, 31(4): 315--319.

\bibitem{fairchild2013color}
Fairchild MD.  \emph{Color appearance models}, John Wiley \& Sons2013.

\bibitem{schade1956optical}
Schade OH. Optical and photoelectric analog of the eye. \emph{JoSA}, 1956,
  46(9): 721--739.

\bibitem{xia1996dynamic}
Xia JC, Varshney A. Dynamic view-dependent simplification for polygonal models.
  In \emph{Proceedings of Seventh Annual IEEE Visualization'96}, IEEE1996,
  327--334.

\bibitem{hoppe1997view}
Hoppe H. View-dependent refinement of progressive meshes. In \emph{Proceedings
  of the 24th annual conference on Computer graphics and interactive
  techniques}, 1997, 189--198.

\bibitem{luebke1997view}
Luebke D, Erikson C. View-dependent simplification of arbitrary polygonal
  environments. In \emph{Proceedings of the 24th annual conference on Computer
  graphics and interactive techniques}, 1997, 199--208.

\bibitem{patney2016towards}
Patney A, Salvi M, Kim J, Kaplanyan A, Wyman C, Benty N, Luebke D, Lefohn A.
  Towards foveated rendering for gaze-tracked virtual reality. \emph{ACM
  Transactions on Graphics (TOG)}, 2016, 35(6): 179.

\bibitem{1984Contrast}
Liu YM, Jiang BC. Contrast sensitivity measured during smooth pursuit movement.
  \emph{Science China Chemistry}, 1984, 27(7): 710.

\bibitem{1988Contrast}
Flipse JP, Wildt GJVD, Rodenburg M, Keemink CJ, Knol P. Contrast sensitivity
  for oscillating sine wave gratings during ocular fixation and pursuit.
  \emph{Vision Research}, 1988, 28(7): 819--826.

\bibitem{1972Resolution}
Meeteren AV, Vos JJ. Resolution and contrast sensitivity at low luminances.
  \emph{Vision Research}, 1972, 12(5): 825,IN2--833,IN2.

\bibitem{1991Colour}
Mullen K. Colour vision as a post-receptoral specialization of the central
  visual field. \emph{Vision Research}, 1991, 31(1): 119--130,
  \doi{https://doi.org/10.1016/0042-6989(91)90079-K}.

\bibitem{anderson1991human}
Anderson SJ, Mullen KT, Hess RF. Human peripheral spatial resolution for
  achromatic and chromatic stimuli: limits imposed by optical and retinal
  factors. \emph{The Journal of Physiology}, 1991, 442(1): 47--64.

\bibitem{2002Differential}
Mullen KT, Kingdom FAA. Differential distributions of red–green and
  blue–yellow cone opponency across the visual field. \emph{Visual
  Neuroscience}, 2002, 19: 109 -- 118.

\bibitem{2005Does}
Mullen KT, Sakurai M, Chu W. Does L/M cone opponency disappear in human
  periphery? \emph{Perception}, 2005, 34(8): 951--959.

\bibitem{duchowski2007foveated}
Duchowski AT, {\c{C}}{\"o}ltekin A. Foveated gaze-contingent displays for
  peripheral LOD management, 3D visualization, and stereo imaging. \emph{ACM
  Transactions on Multimedia Computing, Communications, and Applications
  (TOMM)}, 2007, 3(4): 1--18.

\bibitem{tyler1993eccentricity}
Tyler CW, Hamer RD. Eccentricity and the Ferry--Porter law. \emph{JOSA A},
  1993, 10(9): 2084--2087.

\bibitem{spjut2020toward}
Spjut J, Boudaoud B, Kim J, Greer T, Albert R, Stengel M, Ak{\c{s}}it K, Luebke
  D. Toward Standardized Classification of Foveated Displays. \emph{IEEE
  Transactions on Visualization and Computer Graphics}, 2020, 26(5):
  2126--2134.

\bibitem{duchowski2009spatiochromatic}
Duchowski AT, Bate D, Stringfellow P, Thakur K, Melloy BJ, Gramopadhye AK. On
  spatiochromatic visual sensitivity and peripheral color LOD management.
  \emph{ACM Transactions on Applied Perception (TAP)}, 2009, 6(2): 1--18.

\bibitem{funkhouser1993adaptive}
Funkhouser TA, S{\'e}quin CH. Adaptive display algorithm for interactive frame
  rates during visualization of complex virtual environments. In
  \emph{Proceedings of the 20th annual conference on Computer graphics and
  interactive techniques}, 1993, 247--254.

\bibitem{reddy2001perceptually}
Reddy M. Perceptually optimized 3D graphics. \emph{IEEE computer Graphics and
  Applications}, 2001, 21(5): 68--75.

\bibitem{murphy2001gaze}
Murphy H, Duchowski A. Gaze-Contingent Level Of Detail Rendering.
  \emph{EuroGraphics}, 2001, 2001.

\bibitem{cheng2003foveated}
Cheng I. Foveated 3D model simplification. In \emph{Seventh International
  Symposium on Signal Processing and Its Applications, 2003. Proceedings.},
  volume~1, IEEE2003, 241--244.

\bibitem{duchowski2003gaze}
Duchowski AT, Cournia N, Murphy H. Gaze-Contingent Displays: Review and Current
  Trends, 2003.

\bibitem{reingold2003gaze}
Reingold EM, Loschky LC, McConkie GW, Stampe DM. Gaze-contingent
  multiresolutional displays: An integrative review. \emph{Human factors},
  2003, 45(2): 307--328.

\bibitem{zhou2004distance}
Zhou J, D{\"o}ring A, T{\"o}nnies KD. Distance based enhancement for focal
  region based volume rendering. In \emph{Bildverarbeitung f{\"u}r die Medizin
  2004}, Springer2004, 199--203.

\bibitem{yu2005fast}
Yu H, Chang EC, Huang Z, Zheng Z. Fast rendering of foveated volumes in
  wavelet-based representation. \emph{The Visual Computer}, 2005, 21(8-10):
  735--744.

\bibitem{lu2006volume}
Lu A, Maciejewski R, Ebert DS. Volume composition using eye tracking data. In
  \emph{EuroVis}, 2006, 115--122.

\bibitem{hillaire2008depth}
Hillaire S, L{\'e}cuyer A, Cozot R, Casiez G. Depth-of-field blur effects for
  first-person navigation in virtual environments. \emph{IEEE computer graphics
  and applications}, 2008, 28(6): 47--55.

\bibitem{murphy2009hybrid}
Murphy HA, Duchowski AT, Tyrrell RA. Hybrid Image/Model-Based Gaze-Contingent
  Rendering. \emph{ACM Trans. Appl. Percept.}, 2009, 5(4),
  \doi{10.1145/1462048.1462053}.

\bibitem{gallo2013high}
Gallo L, Placitelli AP. High-fidelity visualization of large medical datasets
  on commodity hardware. \emph{ISRN Biomedical Engineering}, 2013, 2013.

\bibitem{fujita2014foveated}
Fujita M, Harada T. Foveated real-time ray tracing for virtual reality headset.
  \emph{Poster, SIGGRAPH Asia}, 2014, 14.

\bibitem{patney2016perceptually}
Patney A, Kim J, Salvi M, Kaplanyan A, Wyman C, Benty N, Lefohn A, Luebke D.
  Perceptually-Based Foveated Virtual Reality. In \emph{ACM SIGGRAPH 2016
  Emerging Technologies}, SIGGRAPH '16, New York, NY, USA: Association for
  Computing Machinery2016, \doi{10.1145/2929464.2929472}.

\bibitem{swafford2016user}
Swafford NT, Iglesias-Guitian JA, Koniaris C, Moon B, Cosker D, Mitchell K.
  User, metric, and computational evaluation of foveated rendering methods. In
  \emph{Proceedings of the ACM Symposium on Applied Perception}, 2016, 7--14.

\bibitem{pai2016gazesim}
Pai YS, Tag B, Outram B, Vontin N, Sugiura K, Kunze K. GazeSim: simulating
  foveated rendering using depth in eye gaze for VR. In \emph{ACM SIGGRAPH 2016
  Posters}, 2016, 1--2.

\bibitem{lindeberg2016concealing}
Lindeberg T. Concealing rendering simplifications using gazecontingent depth of
  field, 2016.

\bibitem{albert2017latency}
Albert R, Patney A, Luebke D, Kim J. Latency requirements for foveated
  rendering in virtual reality. \emph{ACM Transactions on Applied Perception
  (TAP)}, 2017, 14(4): 1--13.

\bibitem{blackmon2017foveated}
Blackmon S, Peterson LT, Ozdas C, Clohset SJ. Foveated Rendering, 2017, uS
  Patent App. 15/372,589.

\bibitem{koskela2017foveated}
Koskela M, Immonen K, Viitanen T, J{\"a}{\"a}skel{\"a}inen P, Multanen J,
  Takala J. Foveated instant preview for progressive rendering. In
  \emph{SIGGRAPH Asia 2017 Technical Briefs}, 2017, 1--4.

\bibitem{hsu2017foveated}
Hsu CF, Chen A, Hsu CH, Huang CY, Lei CL, Chen KT. Is foveated rendering
  perceivable in virtual reality? Exploring the efficiency and consistency of
  quality assessment methods. In \emph{Proceedings of the 25th ACM
  international conference on multimedia}, 2017, 55--63.

\bibitem{sun2017perceptually}
Sun Q, Huang FC, Kim J, Wei LY, Luebke D, Kaufman A. Perceptually-guided
  foveation for light field displays. \emph{ACM Transactions on Graphics
  (TOG)}, 2017, 36(6): 1--13.

\bibitem{lungaro2018gaze}
Lungaro P, Sj{\"o}berg R, Valero AJF, Mittal A, Tollmar K. Gaze-aware streaming
  solutions for the next generation of mobile VR experiences. \emph{IEEE
  transactions on visualization and computer graphics}, 2018, 24(4):
  1535--1544.

\bibitem{meng2018kernel}
Meng X, Du R, Zwicker M, Varshney A. Kernel foveated rendering.
  \emph{Proceedings of the ACM on Computer Graphics and Interactive
  Techniques}, 2018, 1(1): 1--20.

\bibitem{koskela2018instantaneous}
Koskela MK, Immonen KV, Viitanen TT, J{\"a}{\"a}skel{\"a}inen PO, Multanen JI,
  Takala JH. Instantaneous foveated preview for progressive Monte Carlo
  rendering. \emph{Computational Visual Media}, 2018, 4(3): 267--276.

\bibitem{tan2018foveated}
Tan G, Lee YH, Zhan T, Yang J, Liu S, Zhao D, Wu ST. Foveated imaging for
  near-eye displays. \emph{Optics express}, 2018, 26(19): 25076--25085.

\bibitem{wilson2018rendering}
Wilson A, Lanman DR, Trail ND, McEldowney SC, McNally SJ, Sulai YNB. Rendering
  composite content on a head-mounted display including a high resolution
  inset, 2018, uS Patent 9,972,071.

\bibitem{young2019foveal}
Young A, Ho C, Stafford JR. Foveal adaptation of particles and simulation
  models in a foveated rendering system, 2019, uS Patent 10,339,692.

\bibitem{wei2019fast}
Wei L, Sakamoto Y. Fast calculation method with foveated rendering for
  computer-generated holograms using an angle-changeable ray-tracing method.
  \emph{Applied optics}, 2019, 58(5): A258--A266.

\bibitem{young2019real}
Young A, Stafford JR. Real-time user adaptive foveated rendering, 2019, uS
  Patent 10,192,528.

\bibitem{stafford2019selective}
Stafford JR, Young A. Selective peripheral vision filtering in a foveated
  rendering system, 2019, uS Patent 10,169,846.

\bibitem{friston2019perceptual}
Friston S, Ritschel T, Steed A. Perceptual rasterization for head-mounted
  display image synthesis. \emph{ACM Transactions on Graphics (TOG)}, 2019,
  38(4): 1--14.

\bibitem{radkowski2019impact}
Radkowski R, Raul S. Impact of Foveated Rendering on Procedural Task Training.
  In \emph{International Conference on Human-Computer Interaction},
  Springer2019, 258--267.

\bibitem{siekawa2019foveated}
Siekawa A, Chwesiuk M, Mantiuk R, Pi{\'o}rkowski R. Foveated ray tracing for VR
  headsets. In \emph{International Conference on Multimedia Modeling},
  Springer2019, 106--117.

\bibitem{kim2019foveated}
Kim J, Jeong Y, Stengel M, Ak{\c{s}}it K, Albert R, Boudaoud B, Greer T, Kim J,
  Lopes W, Majercik Z, et~al.. Foveated AR: dynamically-foveated augmented
  reality display. \emph{ACM Transactions on Graphics (TOG)}, 2019, 38(4):
  1--15.

\bibitem{lee2019enhanced}
Lee JS, Kim YK, Lee MY, Won YH. Enhanced see-through near-eye display using
  time-division multiplexing of a Maxwellian-view and holographic display.
  \emph{Optics express}, 2019, 27(2): 689--701.

\bibitem{young2020optimized}
Young A, Ho C, Stafford JR. Optimized shadows in a foveated rendering system,
  2020, uS Patent 10,650,544.

\bibitem{ananpiriyakul2020gaze}
Ananpiriyakul T, Anghel J, Potter K, Joshi A. A gaze-contingent system for
  foveated multiresolution visualization of vector and volumetric data.
  \emph{Electronic Imaging}, 2020, 2020(1): 374--1.

\bibitem{konrad2020gaze}
Konrad R, Angelopoulos A, Wetzstein G. Gaze-contingent ocular parallax
  rendering for virtual reality. \emph{ACM Transactions on Graphics (TOG)},
  2020, 39(2): 1--12.

\bibitem{joshi2020inattentional}
Joshi Y, Poullis C. Inattentional Blindness for Redirected Walking Using
  Dynamic Foveated Rendering. \emph{IEEE Access}, 2020, 8: 39013--39024.

\bibitem{meng20203d}
Meng X, Du R, JaJa JF, Varshney A. 3D-Kernel Foveated Rendering for Light
  Fields. \emph{IEEE Transactions on Visualization and Computer Graphics},
  2020.

\bibitem{friess2020foveated}
Frie{\ss} F, Braun M, Bruder V, Frey S, Reina G, Ertl T. Foveated Encoding for
  Large High-Resolution Displays. \emph{IEEE Transactions on Visualization and
  Computer Graphics}, 2020, 27(2): 1850--1859.

\bibitem{yoo2020foveated}
Yoo C, Xiong J, Moon S, Yoo D, Lee CK, Wu ST, Lee B. Foveated display system
  based on a doublet geometric phase lens. \emph{Optics Express}, 2020, 28(16):
  23690--23702.

\bibitem{bitterli2020spatiotemporal}
Bitterli B, Wyman C, Pharr M, Shirley P, Lefohn A, Jarosz W. Spatiotemporal
  reservoir resampling for real-time ray tracing with dynamic direct lighting.
  \emph{ACM Transactions on Graphics (TOG)}, 2020, 39(4): 148--1.

\bibitem{deza2021emergent}
Deza A, Konkle T. Emergent Properties of Foveated Perceptual Systems, 2021.

\bibitem{yang2021foveated}
Yang Q, Chen Z, Liu Y, Xing G, Zhang Y. Foveated light culling. \emph{Computers
  \& Graphics}, 2021, 97: 200--207,
  \doi{https://doi.org/10.1016/j.cag.2021.04.021}.

\bibitem{franke2021time}
Franke L, Fink L, Martschinke J, Selgrad K, Stamminger M. Time-Warped Foveated
  Rendering for Virtual Reality Headsets. In \emph{Computer Graphics Forum},
  volume~40, Wiley Online Library2021, 110--123.

\bibitem{surace2021learning}
Surace L, Wernikowski M, Tursun O, Myszkowski K, Mantiuk R, Didyk P. Learning
  Foveated Reconstruction to Preserve Perceived Image Statistics. \emph{arXiv
  preprint arXiv:2108.03499}, 2021.

\bibitem{youngwook2021selective}
Youngwook K, Yunmin K, Insung I. Selective Foveated Ray Tracing for
  Head-Mounted Displays. In \emph{2021 IEEE International Symposium on Mixed
  and Augmented Reality (ISMAR)}, IEEE2021, 1--9.

\bibitem{jingyu2021perception}
Jingyu L, Claire M, Søren F. Perception-Driven Hybrid Foveated Depth of Field
  Rendering for Head-Mounted Displays. In \emph{2021 IEEE International
  Symposium on Mixed and Augmented Reality (ISMAR)}, IEEE2021, 1--10.

\bibitem{walton2021beyond}
Walton DR, Anjos RKD, Friston S, Swapp D, Ak{\c{s}}it K, Steed A, Ritschel T.
  Beyond blur: real-time ventral metamers for foveated rendering. \emph{ACM
  Transactions on Graphics (TOG)}, 2021, 40(4): 1--14.

\bibitem{li2021log}
Li D, Du R, Babu A, Brumar CD, Varshney A. A Log-Rectilinear Transformation for
  Foveated 360-degree Video Streaming. \emph{IEEE Transactions on Visualization
  and Computer Graphics}, 2021, 27(5): 2638--2647.

\bibitem{shi2021foveated}
Shi X, Wang L, Wei X, Yan LQ. Foveated Photon Mapping. \emph{IEEE Transactions
  on Visualization and Computer Graphics}, 2021, 27(11): 4183--4193.

\bibitem{chakravarthula2021gaze}
Chakravarthula P, Zhang Z, Tursun O, Didyk P, Sun Q, Fuchs H. Gaze-Contingent
  Retinal Speckle Suppression for Perceptually-Matched Foveated Holographic
  Displays. \emph{IEEE Transactions on Visualization and Computer Graphics},
  2021, 27(11): 4194--4203.

\bibitem{jindal2021perceptual}
Jindal A, Wolski K, Myszkowski K, Mantiuk RK. Perceptual model for adaptive
  local shading and refresh rate. \emph{ACM Transactions on Graphics (TOG)},
  2021, 40(6): 1--18.

\bibitem{alwani2018microsoft}
Alwani R. Microsoft and Nvidia Tech to Bring Photorealistic Games With Ray
  Tracing, 2018.

\bibitem{sanzharov2020survey}
Sanzharov V, Frolov VA, Galaktionov VA. Survey of Nvidia RTX Technology.
  \emph{Programming and Computer Software}, 2020, 46(4): 297--304.

\bibitem{mukhina2015method}
Mukhina K, Bezgodov A. The method for real-time cloud rendering. \emph{Procedia
  Computer Science}, 2015, 66: 697--704.

\bibitem{clark1976hierarchical}
Clark JH. Hierarchical geometric models for visible surface algorithms.
  \emph{Communications of the ACM}, 1976, 19(10): 547--554.

\bibitem{luebke2003level}
Luebke D, Reddy M, Cohen JD, Varshney A, Watson B, Huebner R.  \emph{Level of
  detail for 3D graphics}, Morgan Kaufmann2003.

\bibitem{hoppe1996progressive}
Hoppe H. Progressive meshes. In \emph{Proceedings of the 23rd annual conference
  on Computer graphics and interactive techniques}, 1996, 99--108.

\bibitem{rovamo1979estimation}
Rovamo J, Virsu V. An estimation and application of the human cortical
  magnification factor. \emph{Experimental brain research}, 1979, 37(3):
  495--510.

\bibitem{ramasubramanian1999perceptually}
Ramasubramanian M, Pattanaik SN, Greenberg DP. A perceptually based physical
  error metric for realistic image synthesis. In \emph{Proceedings of the 26th
  annual conference on Computer graphics and interactive techniques}, 1999,
  73--82.

\bibitem{myszkowski2001perception}
Myszkowski K, Tawara T, Akamine H, Seidel HP. Perception-guided global
  illumination solution for animation rendering. In \emph{Proceedings of the
  28th annual conference on Computer graphics and interactive techniques},
  2001, 221--230.

\bibitem{furnas1986generalized}
Furnas GW. Generalized fisheye views. \emph{Acm Sigchi Bulletin}, 1986, 17(4):
  16--23.

\bibitem{sarkar1992graphical}
Sarkar M, Brown MH. Graphical fisheye views of graphs. In \emph{Proceedings of
  the SIGCHI conference on Human factors in computing systems}, 1992, 83--91.

\bibitem{lamping1995focus}
Lamping J, Rao R, Pirolli P. A focus+ context technique based on hyperbolic
  geometry for visualizing large hierarchies. In \emph{Proceedings of the
  SIGCHI conference on Human factors in computing systems}, 1995, 401--408.

\bibitem{carpendale1996distortion}
Carpendale MST, Cowperthwaite DJ, Fracchia FD. Distortion viewing techniques
  for 3-dimensional data. In \emph{Proceedings IEEE Symposium on Information
  Visualization'96}, IEEE1996, 46--53.

\bibitem{plaisant2002spacetree}
Plaisant C, Grosjean J, Bederson BB. Spacetree: Supporting exploration in large
  node link tree, design evolution and empirical evaluation. In \emph{IEEE
  Symposium on Information Visualization, 2002. INFOVIS 2002.}, IEEE2002,
  57--64.

\bibitem{kosara2002focus}
Kosara R, Miksch S, Hauser H. Focus+ context taken literally. \emph{IEEE
  Computer Graphics and Applications}, 2002, 22(1): 22--29.

\bibitem{munzner2003treejuxtaposer}
Munzner T, Guimbreti{\`e}re F, Tasiran S, Zhang L, Zhou Y. TreeJuxtaposer:
  scalable tree comparison using Focus+ Context with guaranteed visibility. In
  \emph{ACM SIGGRAPH 2003 Papers}, 2003, 453--462.

\bibitem{viola2004importance}
Viola I, Kanitsar A, Groller ME. Importance-driven volume rendering. In
  \emph{IEEE Visualization 2004}, IEEE2004, 139--145.

\bibitem{cater2002selective}
Cater K, Chalmers A, Ledda P. Selective quality rendering by exploiting human
  inattentional blindness: looking but not seeing. In \emph{Proceedings of the
  ACM symposium on Virtual reality software and technology}, 2002, 17--24.

\bibitem{cater2003detail}
Cater K, Chalmers A, Ward G. Detail to attention: exploiting visual tasks for
  selective rendering. In \emph{ACM International Conference Proceeding
  Series}, volume~44, 2003, 270--280.

\bibitem{sundstedt2004selective}
Sundstedt V, Chalmers A, Cater K. Selective rendering of task related scenes.
  In \emph{Proceedings of the 1st Symposium on Applied perception in graphics
  and visualization}, 2004, 174--174.

\bibitem{sundstedt2004top}
Sundstedt V, Chalmers A, Cater K, Debattista K. Top-Down Visual Attention for
  Efficient Rendering of Task Related Scenes. In \emph{VMV}, volume~4, 2004,
  209--216.

\bibitem{duchowski1995simple}
Duchowski AT, McCormick BH. Simple multiresolution approach for representing
  multiple regions of interest (ROIs). In \emph{Visual Communications and Image
  Processing'95}, volume 2501, International Society for Optics and
  Photonics1995, 175--186.

\bibitem{geisler1999variable}
Geisler WS, Perry JS. Variable-Resolution Displays for Visual Communication and
  Simulation. In \emph{SID Symposium Digest of Technical Papers}, volume~30,
  Wiley Online Library1999, 420--423.

\bibitem{parkhurst2000evaluating}
Parkhurst D, Culurciello E, Niebur E. Evaluating variable resolution displays
  with visual search: Task performance and eye movements. In \emph{Proceedings
  of the 2000 symposium on Eye tracking research \& applications}, 2000,
  105--109.

\bibitem{geisler2002real}
Geisler WS, Perry JS. Real-time simulation of arbitrary visual fields. In
  \emph{Proceedings of the 2002 symposium on Eye tracking research \&
  applications}, 2002, 83--87.

\bibitem{willberger2019deferred}
Willberger T, Musterle C, Bergmann S. Deferred hybrid path tracing. In
  \emph{Ray Tracing Gems}, Springer2019, 475--492.

\bibitem{jin2009selective}
Jin B, Ihm I, Chang B, Park C, Lee W, Jung S. Selective and Adaptive
  Supersampling for Real-Time Ray Tracing. In \emph{Proceedings of the
  Conference on High Performance Graphics 2009}, HPG '09, New York, NY, USA:
  Association for Computing Machinery2009, 117–125,
  \doi{10.1145/1572769.1572788}.

\bibitem{koskela2019Blockwise}
Koskela M, Immonen K, Mäkitalo M, Foi A, Takala J. Blockwise Multi-Order
  Feature Regression for Real-Time Path-Tracing Reconstruction. \emph{ACM
  Transactions on Graphics (TOG)}, 2019, 38(5): 1--14.

\bibitem{sherrington1897onReciprocal}
Sherrington CS. On Reciprocal Action in the Retina as studied by means of some
  Rotating Discs. \emph{The Journal of Physiology}, 1897, 21(1): 33--54,
  \doi{https://doi.org/10.1113/jphysiol.1897.sp000641}.

\bibitem{arabadzhiyska2017saccade}
Arabadzhiyska E, Tursun OT, Myszkowski K, Seidel HP, Didyk P. Saccade landing
  position prediction for gaze-contingent rendering. \emph{ACM Transactions on
  Graphics (TOG)}, 2017, 36(4): 1--12.

\bibitem{mlot20163d}
Mlot EG, Bahmani H, Wahl S, Kasneci E. 3D Gaze Estimation using Eye Vergence.
  In \emph{HEALTHINF}, 2016, 125--131.

\bibitem{veach1997robust}
Veach E. \emph{Robust Monte Carlo methods for light transport simulation},
  volume 1610, Stanford University PhD thesis1997.

\bibitem{georgiev2012light}
Georgiev I, Kriv{\'a}nek J, Davidovic T, Slusallek P. Light transport
  simulation with vertex connection and merging. \emph{ACM Trans. Graph.},
  2012, 31(6): 192--1.

\bibitem{lee2002foveated}
Lee S, Pattichis MS, Bovik AC. Foveated video quality assessment. \emph{IEEE
  Transactions on Multimedia}, 2002, 4(1): 129--132.

\bibitem{wang2003foveated}
Wang Z, Bovik A, Lu L. Foveated Wavelet Image Quality Index. \emph{Proceedings
  of SPIE - The International Society for Optical Engineering}, 2003, 4472,
  \doi{10.1117/12.449797}.

\bibitem{you2013attention}
You J, Ebrahimi T, Perkis A. Attention driven foveated video quality
  assessment. \emph{IEEE Transactions on Image Processing}, 2013, 23(1):
  200--213.

\bibitem{rimac2010foveated}
Rimac-Drlje S, Vranje{\v{s}} M, {\v{Z}}agar D. Foveated mean squared error—a
  novel video quality metric. \emph{Multimedia tools and applications}, 2010,
  49(3): 425--445.

\bibitem{tsai2014foveation}
Tsai WJ, Liu YS. Foveation-based image quality assessment. In \emph{2014 IEEE
  Visual Communications and Image Processing Conference}, IEEE2014, 25--28.

\end{thebibliography}

\subsection*{Author biography}

\begin{biography}[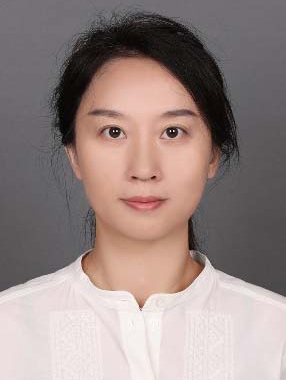]{Lili Wang} is a professor at the School of Computer Science and Engineering, Beihang University, and a researcher of the State Key Laboratory of Virtual Reality Technology and Systems. Her research interests include virtual reality, augmented reality, and rendering.
\end{biography}

\begin{biography}[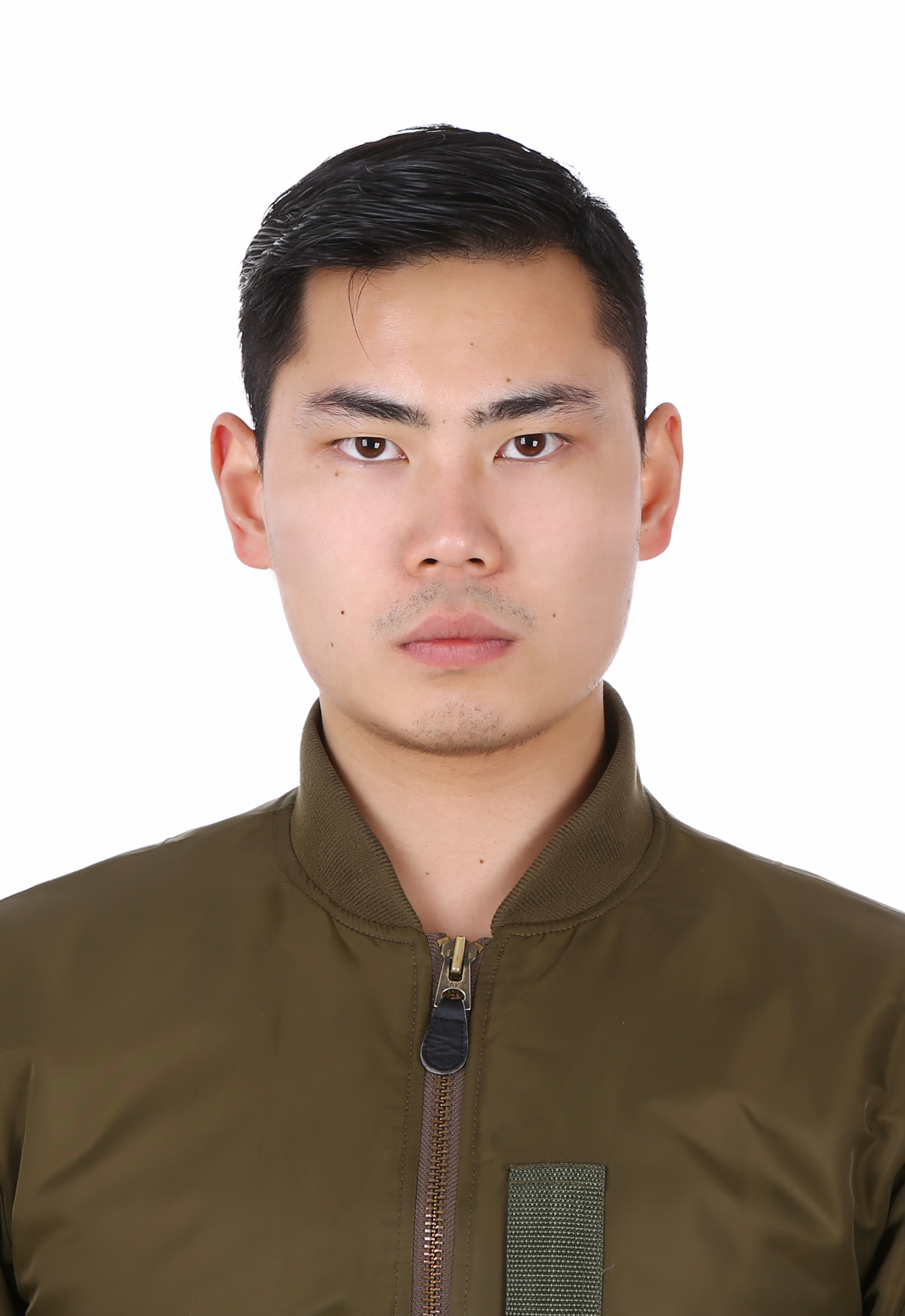]{Xuehuai Shi} is a Ph.D student at the School of Computer Science and Engineering, Beihang University, China. His current research focuses on virtual reality and foveated rendering.
\end{biography}

\begin{biography}[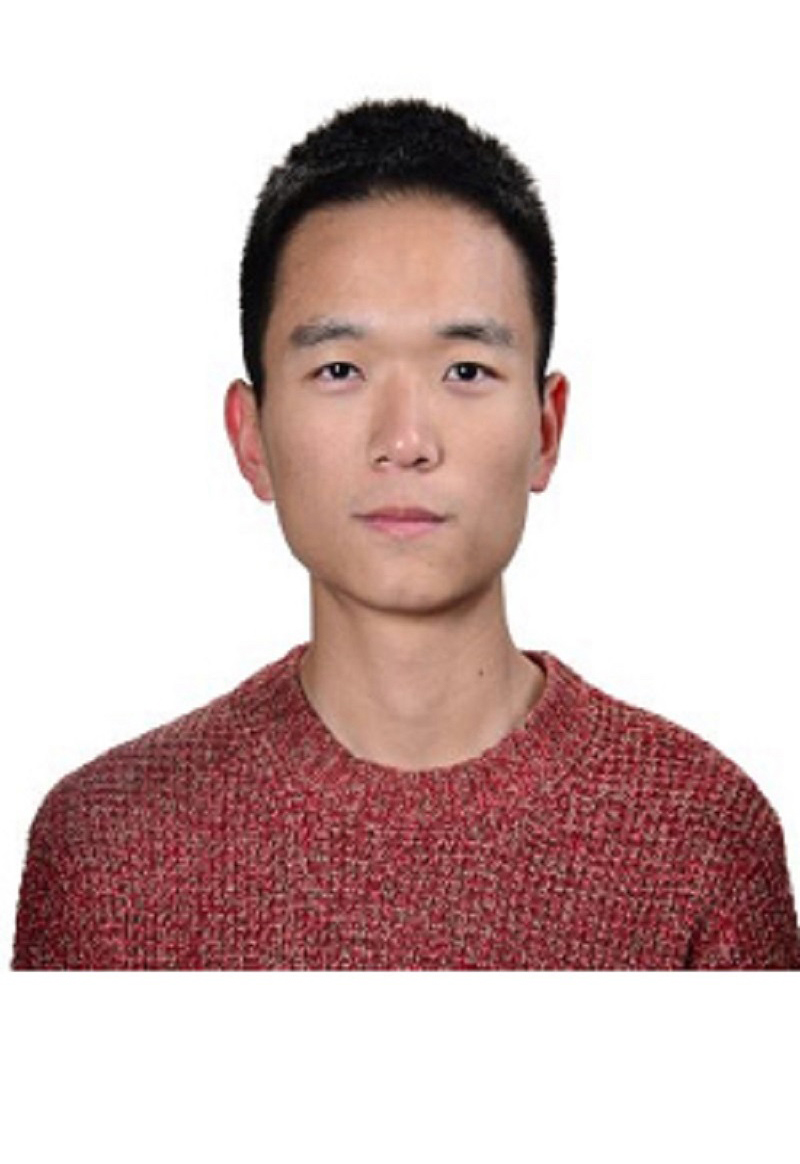]{Yi Liu} is a master student at the School of Computer Science and Engineering, Beihang University, China. His current research focuses on virutal reailty, augumented reality and rendering.
\end{biography}
%

\end{document}